\documentclass[prd,twocolumn,floatfix,amsmath,nofootinbib,amssymb,floatfix]{revtex4}
\usepackage{graphicx,color,dcolumn,booktabs,bm}
\usepackage{longtable,lscape}
\usepackage{pdfpages}
\usepackage{txfonts}
\usepackage{overpic}
\usepackage{amssymb}
\usepackage{makecell}
\usepackage{indentfirst}
\usepackage{feynmf}   
\usepackage{slashed}  
\usepackage{cases}
\usepackage{color}
\usepackage{multirow}
\usepackage{threeparttable}
\usepackage{epstopdf}
\usepackage{enumerate}
\usepackage{subfigure}
\usepackage{textcomp}
\usepackage{diagbox}
\usepackage{graphicx,color,dcolumn,booktabs,bm}
\usepackage{mathrsfs}
\usepackage{cancel}
\usepackage{float}
\usepackage[misc]{ifsym}
\usepackage[colorlinks,
citecolor=blue,
anchorcolor=red,
menucolor=red,
linkcolor=red,
filecolor=red,
runcolor=red,
urlcolor=blue,
frenchlinks=red]{hyperref}
\usepackage{array}
\usepackage{booktabs}

\begin{document}
\title{Analysis of the semileptonic decays of $\Xi_{cc}$ and $\Omega_{cc}$ baryons in QCD sum rules}
\author{Guo-Liang Yu$^{1,3}$}
\email{yuguoliang2011@163.com}
\author{Zhi-Gang Wang$^{1,3}$}
\email{zgwang@aliyun.com}
\author{Jie Lu$^{2}$}
\email{l17693567997@163.com}
\author{Bin Wu$^{2}$}
\author{Peng Yang$^{1}$}
\author{Ze Zhou$^{1}$}

\affiliation{$^1$ Department of Mathematics and Physics, North China
Electric Power University, Baoding 071003, People's Republic of
China\\$^2$ School of Physics, Southeast University, Nanjing 210094, People's Republic of
China\\$^3$ Hebei Key Laboratory of Physics and Energy Technology,
North China Electric Power University, Baoding 071000, China}
\date{\today }

\begin{abstract}
We firstly carry out a systematic analysis on the spin $\frac{1}{2}^{+}\rightarrow\frac{3}{2}^{+}$ weak transition process in the framework of three-point QCD sum rules, where the initial and final states are doubly and singly charmed baryons. In the phenomenological side, all possible couplings of interpolating current to hadronic states are considered. In doing operator production expansion at QCD side, the contributions of the perturbative part, vacuum condensate terms of $\langle{\bar qq}\rangle$, $\langle g_{s}^{2}GG\rangle$, $\langle \bar q g_{s}\sigma Gq\rangle$ and $g_{s}^{2}\langle{\bar qq}\rangle^{2}$ are all considered. After the form factors in space-like region ($Q^2>0$) are obtained, the numerical results are extrapolated into time-like region ($Q^2<0$) by a fitting function. Using the predicted form factors, we finally analyze the semileptonic decays of $\Xi_{cc}^{++}\rightarrow \Sigma_{c}^{*+}l^{+}\nu_{l}$, $\Xi_{cc}^{++}\rightarrow \Xi_{c}^{\prime*+}l^{+}\nu_{l}$, $\Omega_{cc}^{+}\rightarrow\Xi_{c}^{\prime*0}l^{+}\nu_{l}$ and $\Omega_{cc}^{+}\rightarrow \Omega_{c}^{*0}l^{+}\nu_{l}$ with $l=e,\mu$. The predictions in this work can deepen our understanding of the dynamics in the decay processes of doubly heavy baryons and provide useful information to explore the possibility of new physics in heavy baryonic decay channels.
\end{abstract}


\maketitle

\section{Introduction}\label{sec1}

In recent years, many single heavy baryons have been discovered by Belle, BABAR, CLEO and LHCb collaborations \cite{ParticleDataGroup:2024cfk}, which is like the bamboo shoots after a spring rain. However, searching for doubly heavy baryons in experiments ended with no progress for a long time. The breakthrough came in 2017 with the discovery of the doubly charmed baryon $\Xi_{cc}^{++}$ by the LHCb collaboration \cite{352117}. This state was first observed in the decay channel $\Xi_{cc}^{++}\rightarrow \Lambda_{c}^{+}K^{-}\pi^{+}\pi^{+}$ with a measured mass $3621.40\pm0.72\pm0.14\pm0.27$ MeV and later was confirmed in another decay mode $\Xi_{cc}^{++}\rightarrow \Lambda_{c}^{+}\pi^{+}$ \cite{352118,352119}. We believe that more and more doubly heavy baryons will be discovered with the efforts of experimental and theoretical physicists. In experiments, the firstly-discovered new type of baryons are usually the ground states which can only be reconstructed via weak decay final states. Thus, studying the decay properties of doubly heavy baryons, especially the weak decaying processes, is of great interest to theoretical physicist, which can provide important information to experimentalists for searching for new doubly heavy baryons.

As for doubly charmed baryons, key transitions of interest include the semileptonic decays, such as $\Xi_{cc}^{++}\rightarrow \Xi_{c}^{\prime*+}l^{+}\nu_{l}$, where form factors parameterize the hadronic matrix elements of the weak current. Thus, a central aspect of understanding the structure and dynamics of the doubly charmed baryons lies in the calculation of their form factors. In addition, precise knowledge of these form factors is essential for extracting fundamental parameters, notably the Cabibbo-Kobayashi-Maskawa (CKM) matrix elements like $V_{cs}$/$V_{cd}$ which is not only an important parameter to deepen our understanding to the dynamics of the weak transition, but also is the crucial step of searching for new physics beyond the Standard Model (SM).

Although QCD has achieved remarkable success in the high energy regions, it has non-perturbative properties at low energy regions which makes it difficult to understand the interactions in hadronic level, especially in systems involving heavy quarks. In calculating the form factors and studying weak decays of heavy baryons, physicists mainly employed some phenomenological methods such as the quark models \cite{Faustov:2018ahb,Geng:2020ofy,Albertus:2004wj,Ebert:2006rp,Cheng:1996cs,Ivanov:1997hi,Ivanov:1997ra,Faessler:2009xn,Gutsche:2018utw,Gutsche:2019iac}, the flavor symmetry method \cite{Wang:2017azm} and the light-front quark model \cite{Zhao:2018mrg,Zhao:2018zcb,Chua:2018lfa,Chua:2019yqh,Ke:2019smy,Zhu:2018jet,Li:2021qod,Hu:2020mxk,Li:2021kfb,Lu:2023rmq}, to carry out this research. As for quark model, the baryons are usually reduced to be a quark$-$diquark system and the final results of this method depend strongly on the space functions of the initial and final baryons. In addition, most of these above research works concentrate on spin $\frac{1}{2}\rightarrow\frac{1}{2}$ transition process, few works focus on $\frac{1}{2}\rightarrow\frac{3}{2}$ process with the initial and final state being doubly heavy baryon and singly one, respectively. Thus, it is an interesting and important work to perform a systematic analysis about this weak decay process. In this field, QCD sum rules (QCDSR) is a very effective non-perturbative method, and it has been widely used to investigate the properties of hadrons including the mass spectra, pole residues, strong coupling constants~\cite{Shifman:1978bx,Shifman:1978by,Colangelo:2000dp,Wang:2025sic,Wang:2018lhz,Wang:2017vtv,Zhang:2025qmg,Lu:2025zaf}, and the form factors \cite{Wang:2008sm,Wang:2015ndk,Khodjamirian:2011jp,Zhao:2020mod,Shi:2019hbf,Azizi:2011mw,Aliev:2023tpk,Khajouei:2024frw,Miao:2022bga,
Xing:2021enr,Zhao:2021sje,Zhang:2023nxl,Neishabouri:2024gbc,Tousi:2024usi}.

In the frame work of three-point QCDSR, it will be seen in the following section (Sec. \ref{sec2}) that there are many different dirac structures both at the phenomenological and QCD sides. The form factors $F_{i}$ and $G_{i}$ ($i=1\sim4$) for the transition process of  $\frac{1}{2}\rightarrow\frac{3}{2}$ are not independent of each other and each of them has relations with several structures. That is to say, each form factor can not be determined only according to one dirac structure. In our previous work \cite{Lu:2025gol}, we successfully analyzed the form factors of $\Lambda_{b}\rightarrow\Lambda_{c}$ and $\Xi_{b}\rightarrow\Xi_{c}$ by solving 12 linear equations about the form factors, where all the dirac structures are considered and the vacuum condensates up to dimension 8 are included at the QCD side. In the present work, we will analyze the form factors of transition processes $\Xi_{cc}^{++}\rightarrow\Sigma_{c}^{*+}$, $\Xi_{cc}^{++}\rightarrow\Xi_{c}^{\prime*+}$, $\Omega_{cc}^{+}\rightarrow\Xi_{c}^{\prime*0}$ and $\Omega_{cc}^{+}\rightarrow\Omega_{c}^{*0}$. As an application of the form factors, the corresponding semileptonic decay processes will also be analyzed.

This article is organized as follows. In Sec. \ref{sec2}, we introduce how the form factors of transition processes $\Xi_{cc}^{++}\rightarrow\Sigma_{c}^{*+}$, $\Xi_{cc}^{++}\rightarrow\Xi_{c}^{\prime*+}$, $\Omega_{cc}^{+}\rightarrow\Xi_{c}^{\prime*0}$ and $\Omega_{cc}^{+}\rightarrow\Omega_{c}^{*0}$ are analyzed in the frame work of QCDSR. In Sec. \ref{sec3}, we introduce in detail how the parameters of QCDSR in the present work are determined, and compare our numerical results with those of other collaborations. With these form factors, the semileptonic decays with spin-parity $\frac{1}{2}^{+}\rightarrow\frac{3}{2}^{+}$ are analyzed in Sec. \ref{sec4}. Sec. \ref{sec5} is reserved for the conclusion part. Some complicated formulas are shown in Appendices~\ref{Sec:AppA} and \ref{Sec:AppB}.

\section{QCD sum rules for the form factors}\label{sec2}
\subsection{Transition form factors of $\frac{1}{2}^{+}\rightarrow\frac{3}{2}^{+}$} \label{sec2.1}

\begin{figure}
	\centering
	\includegraphics[width=5.5cm]{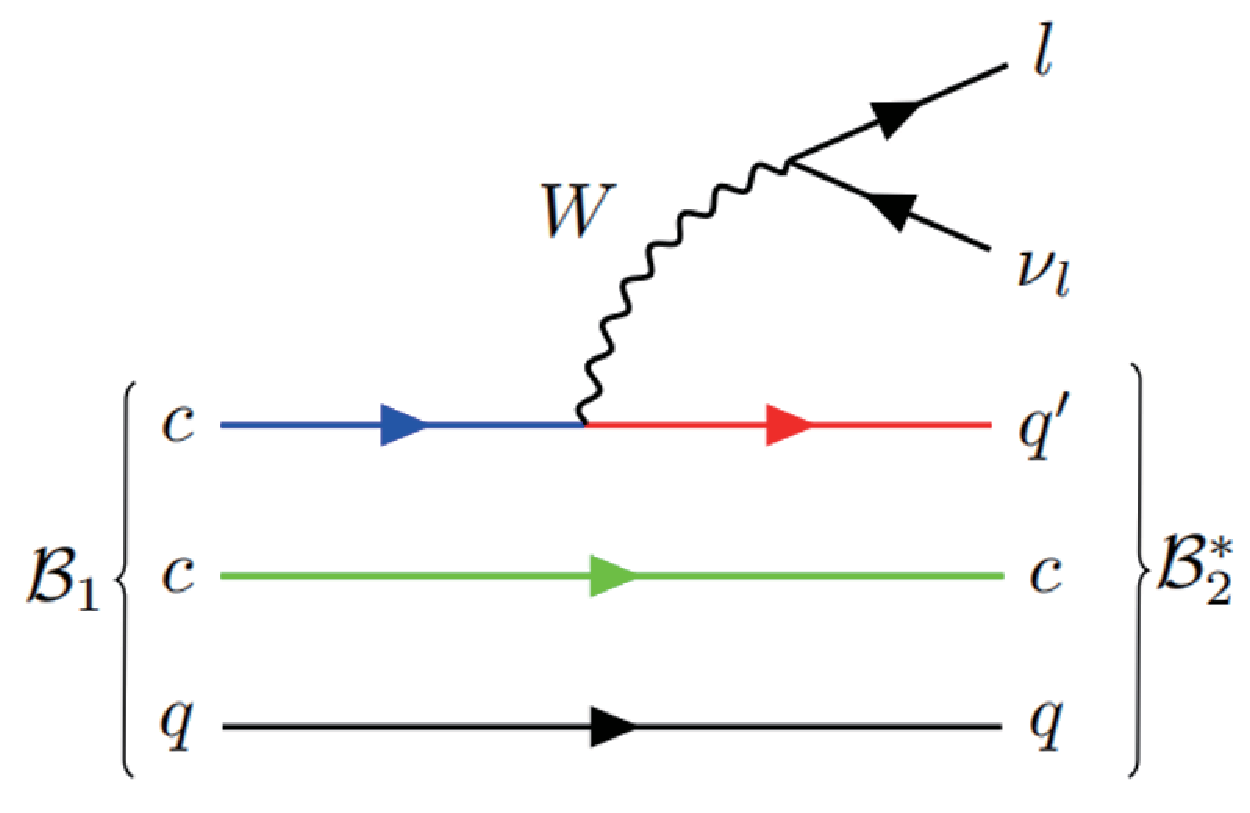}
	\caption{The Feynman diagram of semi-leptonic decay processes of doubly charmed baryons.}
	\label{Fig1}
\end{figure}

It is known that the transition form factors are key parameters in analyzing the semileptonic decay processes. The feynman diagram of this process can be explicitly illustrated in Fig. \ref{Fig1}, where it includes the transition $c\to q^{\prime}l\bar{\nu}_l$ at the quark level with $q^{\prime}=d,s$. In this transition, the leptonic amplitude can be calculated using electro-weak perturbation theory, while the hadronic transition matrix element can not be calculated perturbatively due to non-perturbation effect of low energy QCD. Commonly, it can be parameterized into several transition form factors as in Eq. \ref{eq:1}. In this equation, $J^{V-A}_{\nu}$ is the electroweak transition current with $J^{V-A}_{\nu}= \overline{q}^{\prime}\gamma_{\nu}(1-\gamma_{5})c$ which include the vector and axial-vector two parts. $u(p,s)$ and $u_{\alpha}(p^{\prime},s^{\prime})$ denote the spinor wave functions of initial ($\mathcal{B}_{1}$) and final ($\mathcal{B}_{2}^{*}$) baryons. $F_i(q^2)$ and $G_i(q^2)$ $(i=1\sim4)$ are the vector and axial vector form factors with $q=p-p'$.
\begin{widetext}
\begin{eqnarray}\label{eq:1}
\notag
\left\langle \mathcal{B}_{2}^{*}\left(p^{\prime}\right)|J^{V-A}_{\nu}|\mathcal{B}_{1}\left(p\right) \right\rangle&&=\overline{u}_{\alpha}^{\mathcal{B}_{2}^{*}}\left(p^{\prime},s^{\prime}\right)\gamma_{5}\left[\gamma_{\nu}\frac{p_{\alpha}}{m_{\mathcal{B}_{1}}}F_{1}\left(q^{2}\right)
+\frac{p_{\alpha}p_{\nu}}{m_{\mathcal{B}_{1}}^{2}}F_{2}\left(q^{2}\right)+\frac{p_{\alpha}p^{\prime}_{\nu}}{m_{\mathcal{B}_{1}}m_{\mathcal{B}_{2}^{*}}}F_{3}\left(q^{2}\right)+g_{\alpha\nu}F_{4}\left(q^{2}\right)\right]u^{\mathcal{B}_{1}}\left(p,s\right) \\
&&-\overline{u}_{\alpha}^{\mathcal{B}_{2}^{*}}\left(p^{\prime},s^{\prime}\right)\left[\gamma_{\nu}\frac{p_{\alpha}}{m_{\mathcal{B}_{1}}}G_{1}\left(q^{2}\right)
+\frac{p_{\alpha}p_{\nu}}{m_{\mathcal{B}_{1}}^{2}}G_{2}\left(q^{2}\right)+\frac{p_{\alpha}p^{\prime}_{\nu}}{m_{\mathcal{B}_{1}}m_{\mathcal{B}_{2}^{*}}}G_{3}\left(q^{2}\right)+g_{\alpha\nu}G_{4}\left(q^{2}\right)\right]u^{\mathcal{B}_{1}}\left(p,s\right)
\end{eqnarray}
\end{widetext}

\subsection{QCD sum rules}\label{sec2.2}
To study the form factors of $\Xi_{cc}\rightarrow\Sigma_{c}^{*}$, $\Xi_{cc}\rightarrow\Xi_{c}^{\prime*}$, $\Omega_{cc}\rightarrow\Xi_{c}^{\prime*}$ and $\Omega_{cc}\rightarrow\Omega_{c}^{*}$, we firstly write the following three-point correlation function,
\begin{eqnarray}\label{eq:2}
\notag
\Pi _{\mu\nu}^{\mathrm{V/A}} (p,p') &&= i^2\int d^4x d^4ye^{ip'x}e^{i(p-p')y} \\
&&\times \left\langle 0 \right|T\{J_{\mu}^{\mathcal{B}_{2}^{*}}(x)J^{V-A}_{\nu}(y)\bar J^{\mathcal{B}_{1}}(0)\} \left| 0 \right\rangle
\end{eqnarray}
where $T$ denotes the time ordered product and $\bar{J}=J^\dagger\gamma_0$. $J^{\mathcal{B}_{1}}$ and $J^{\mathcal{B}^{*}_{2}}$ are the interpolating currents of initial and final baryons, respectively. These baryonic currents are taken as the following forms,
\begin{eqnarray}\label{eq:3}
\notag
&&J^{\Xi _{cc}}= \varepsilon _{nml}\left({c^{nT}\mathcal{C}\gamma _{\mu}c^m}\right)\gamma _{\mu}\gamma _{5}q^l \\
\notag
&&J^{\Omega _{cc}}= \varepsilon _{nml}\left({c^{nT}\mathcal{C}\gamma _{\mu}c^m}\right)\gamma _{\mu}\gamma _{5}s^l \\
\notag
&&J^{\Sigma _{c}^{*}}_{\mu}= \frac{1}{\sqrt{2}}\varepsilon _{nml}\left({u^{nT}\mathcal{C}\gamma _{\mu}d^m}+{d^{nT}\mathcal{C}\gamma _{\mu}u^m}\right)c^l \\
\notag
&&J^{\Xi _{c}^{\prime*}}_{\mu}= \frac{1}{\sqrt{2}}\varepsilon _{nml}\left({u^{nT}\mathcal{C}\gamma _{\mu}s^m}+{s^{nT}\mathcal{C}\gamma _{\mu}u^m}\right)c^l \\
&&J^{\Omega _{c}^{*}}_{\mu}= \varepsilon _{nml}\left({s^{nT}\mathcal{C}\gamma _{\mu}s^m}\right)c^l
\end{eqnarray}
where $\varepsilon_{nml}$ is the 3 dimension Levi-Civita tensor, $n$, $m$ and $l$ represent the color indices, and $\mathcal{C}$ is the charge conjugation operator. In the framework of QCDSR, the above three-point correlation function Eq. (\ref{eq:2}) will be calculated at both hadron and quark levels, which are called the phenomenological side and the QCD side, respectively.

In the phenomenological side, a complete sets of hadron states which can couple to the corresponding interpolating currents are inserted into the correlation function. Finishing the integral of coordinate space and using the double dispersion relation~\cite{Shifman:1978bx,Shifman:1978by}, we write the three-point correlation function as,
\begin{eqnarray}\label{eq:4}
\notag
&&{\Pi _{\mu\nu}^{\mathrm{V/A}} }(p,p') \\
\notag
= &&\frac{{\left\langle 0 \right|J_{\mu}^{\mathcal{B}^{*}_{2}}\left| {\mathcal{B}^{*+}_{2}(p')} \right\rangle  \left\langle {\mathcal{B}^{*+}_{2}(p')} \right|J^{V-A}_{\nu}\left| {\mathcal{B}^{+}_{1}(p)} \right\rangle \left\langle {\mathcal{B}^{+}_{1}(p)} \right|{{\bar J}^{\mathcal{B}_{1}}}\left| 0 \right\rangle }}{{(p{'^2} - m_{{\mathcal{B}^{*+}_{2}}}^2)({p^2} - m_{\mathcal{B}_{1}^{+}}^2)}}\\
\notag
&&+ \frac{{\left\langle 0 \right|J_{\mu}^{\mathcal{B}^{*}_{2}}\left| {\mathcal{B}^{*-}_{2}(p')} \right\rangle  \left\langle {\mathcal{B}^{*-}_{2}(p')} \right|J^{V-A}_{\nu}\left| {\mathcal{B}^{+}_{1}(p)} \right\rangle \left\langle {\mathcal{B}^{+}_{1}(p)} \right|{{\bar J}^{\mathcal{B}_{1}}}\left| 0 \right\rangle }}{{(p{'^2} - m_{{\mathcal{B}^{*-}_{2}}}^2)({p^2} - m_{\mathcal{B}_{1}^{+}}^2)}}\\
\notag
&&+ \frac{{\left\langle 0 \right|J_{\mu}^{\mathcal{B}^{*}_{2}}\left| {\mathcal{B}^{+}_{2}(p')} \right\rangle  \left\langle {\mathcal{B}^{+}_{2}(p')} \right|J^{V-A}_{\nu}\left| {\mathcal{B}^{+}_{1}(p)} \right\rangle \left\langle {\mathcal{B}^{+}_{1}(p)} \right|{{\bar J}^{\mathcal{B}_{1}}}\left| 0 \right\rangle }}{{(p{'^2} - m_{{\mathcal{B}^{+}_{2}}}^2)({p^2} - m_{\mathcal{B}_{1}^{+}}^2)}}\\
\notag
&&+ \frac{{\left\langle 0 \right|J_{\mu}^{\mathcal{B}^{*}_{2}}\left| {\mathcal{B}^{-}_{2}(p')} \right\rangle  \left\langle {\mathcal{B}^{-}_{2}(p')} \right|J^{V-A}_{\nu}\left| {\mathcal{B}^{+}_{1}(p)} \right\rangle \left\langle {\mathcal{B}^{+}_{1}(p)} \right|{{\bar J}^{\mathcal{B}_{1}}}\left| 0 \right\rangle }}{{(p{'^2} - m_{{\mathcal{B}^{-}_{2}}}^2)({p^2} - m_{\mathcal{B}_{1}^{+}}^2)}}\\
\notag
&&+\frac{{\left\langle 0 \right|J_{\mu}^{\mathcal{B}^{*}_{2}}\left| {\mathcal{B}^{*+}_{2}(p')} \right\rangle  \left\langle {\mathcal{B}^{*+}_{2}(p')} \right|J^{V-A}_{\nu}\left| {\mathcal{B}^{-}_{1}(p)} \right\rangle \left\langle {\mathcal{B}^{-}_{1}(p)} \right|{{\bar J}^{\mathcal{B}_{1}}}\left| 0 \right\rangle }}{{(p{'^2} - m_{{\mathcal{B}^{*+}_{2}}}^2)({p^2} - m_{\mathcal{B}_{1}^{-}}^2)}}\\
\notag
&&+ \frac{{\left\langle 0 \right|J_{\mu}^{\mathcal{B}^{*}_{2}}\left| {\mathcal{B}^{*-}_{2}(p')} \right\rangle  \left\langle {\mathcal{B}^{*-}_{2}(p')} \right|J^{V-A}_{\nu}\left| {\mathcal{B}^{-}_{1}(p)} \right\rangle \left\langle {\mathcal{B}^{-}_{1}(p)} \right|{{\bar J}^{\mathcal{B}_{1}}}\left| 0 \right\rangle }}{{(p{'^2} - m_{{\mathcal{B}^{*-}_{2}}}^2)({p^2} - m_{\mathcal{B}_{1}^{-}}^2)}}\\
\notag
&&+ \frac{{\left\langle 0 \right|J_{\mu}^{\mathcal{B}^{*}_{2}}\left| {\mathcal{B}^{+}_{2}(p')} \right\rangle  \left\langle {\mathcal{B}^{+}_{2}(p')} \right|J^{V-A}_{\nu}\left| {\mathcal{B}^{-}_{1}(p)} \right\rangle \left\langle {\mathcal{B}^{-}_{1}(p)} \right|{{\bar J}^{\mathcal{B}_{1}}}\left| 0 \right\rangle }}{{(p{'^2} - m_{{\mathcal{B}^{+}_{2}}}^2)({p^2} - m_{\mathcal{B}_{1}^{-}}^2)}}\\
\notag
&&+ \frac{{\left\langle 0 \right|J_{\mu}^{\mathcal{B}^{*}_{2}}\left| {\mathcal{B}^{-}_{2}(p')} \right\rangle  \left\langle {\mathcal{B}^{-}_{2}(p')} \right|J^{V-A}_{\nu}\left| {\mathcal{B}^{-}_{1}(p)} \right\rangle \left\langle {\mathcal{B}^{-}_{1}(p)} \right|{{\bar J}^{\mathcal{B}_{1}}}\left| 0 \right\rangle }}{{(p{'^2} - m_{{\mathcal{B}^{-}_{2}}}^2)({p^2} - m_{\mathcal{B}_{1}^{-}}^2)}} \\
&&+\cdots
\end{eqnarray}
where the ellipsis denote the contributions of excited and continuum states, $\mathcal{B}^{+(-)}_{1}$ and $\mathcal{B}^{*+(-)}_{2}$ represent the initial and final positive (negative) parity bayons. The hadron vacuum matrix elements for initial and final baryons are defined as,
\begin{eqnarray}\label{eq:5}
\notag
&&\left\langle 0 \right|J_{\mu}^{\mathcal{B}_{2}^{*}}(0)\left| \mathcal{B}_{2}^{*+}(p^{\prime},s^{\prime}) \right\rangle  = \lambda_{\mathcal{B}_{2}^{*+}}u_{\mu}(p^{\prime},s^{\prime})\\
\notag
&&\left\langle 0 \right|J_{\mu}^{\mathcal{B}_{2}^{*}}(0)\left| \mathcal{B}_{2}^{*-}(p^{\prime},s^{\prime}) \right\rangle  = \lambda_{\mathcal{B}_{2}^{*-}}i\gamma_{5}u_{\mu}(p^{\prime},s^{\prime})\\
\notag
&&\left\langle 0 \right|J_{\mu}^{\mathcal{B}_{2}^{*}}(0)\left| \mathcal{B}_{2}^{+}(p^{\prime},s^{\prime}) \right\rangle  = \lambda_{\mathcal{B}_{2}^{+}}i\gamma_{5}\Big(\alpha\gamma_{\mu}-\frac{4\alpha}{m}p^{\prime}_{\mu}\Big)u(p^{\prime},s^{\prime})\\
\notag
&&\left\langle 0 \right|J_{\mu}^{\mathcal{B}_{2}^{*}}(0)\left| \mathcal{B}_{2}^{-}(p^{\prime},s^{\prime}) \right\rangle  = \lambda_{\mathcal{B}_{2}^{-}}\Big(\alpha\gamma_{\mu}-\frac{4\alpha}{m}p^{\prime}_{\mu}\Big)u(p^{\prime},s^{\prime})\\ \notag
&&\left\langle B_{1}^{+}(p,s)\right|\overline{J}^{B_{1}}\left|0\right\rangle=\lambda_{B_{1}^{+}}\overline{u}(p,s)\\
&&\left\langle B_{1}^{-}(p,s)\right|\overline{J}^{B_{1}}\left|0\right\rangle=\lambda_{B_{1}^{-}}\overline{u}(p,s)i\gamma_{5}
\end{eqnarray}
Here, $u(p,s)$ and $u_{\mu}(p^{\prime},s^{\prime})$ are the spinor wave functions of $J^{P}=\frac{1}{2}^{\pm}$ and $\frac{3}{2}^{\pm}$ baryons which satisfy the following spin polarization summation formula,
\begin{eqnarray}\label{eq:6}
\sum\limits_{s} u(p,s)\bar u(p,s) = \slashed p + m_{\mathcal{B}_{1}}
\end{eqnarray}
\begin{eqnarray}\label{eq:7}
\notag
&&\sum\limits_{s^{\prime}}u_{\mu}^{\mathcal{B}_{2}^{*}}\left(p^{\prime},s^{\prime}\right)\overline{u}_{\alpha}^{\mathcal{B}_{2}^{*}}\left(p^{\prime},s^{\prime}\right)
=-\left(\slashed p^{\prime}+m_{\mathcal{B}_{2}^{*}}\right)\\
&&\times\Big(g_{\mu\alpha}-\frac{\gamma_{\mu}\gamma_{\alpha}}{3}-\frac{2}{3}\frac{p^{\prime}_{\mu}p^{\prime}_{\alpha}}{m^{2}_{\mathcal{B}_{2}^{*}}}
+\frac{1}{3}\frac{p^{\prime}_{\mu}\gamma_{\alpha}-p^{\prime}_{\alpha}\gamma_{\mu}}{m_{\mathcal{B}_{2}^{*}}}\Big)
\end{eqnarray}
As for the hadronic transition elements $\left\langle {\mathcal{B}^{(*)\pm}_{2}} \right|J^{V-A}_{\nu}\left| {\mathcal{B}^{\pm}_{1}} \right\rangle$, they can be parameterized as a series of form factors which are illustrated in Eq. (\ref{eq:A1}) in Appendix \ref{Sec:AppA}. In this equation, $F^{+-}_{i}$/$G^{+-}_{i}$ ($i=1\sim4$) are the form factors with the positive-parity initial state and the negative-parity final state, and son on. By substituting the matrix elements in Eq. (\ref{eq:4}) with Eqs. (\ref{eq:5})
and (\ref{eq:A1}), the correlation function in the phenomenological side can be decomposed into different dirac structures. From Eq. (\ref{eq:4}), we can see that the current $J^{\mathcal{B}_{1}}$ can couple to the baryons with $J^{P}=\frac{1}{2}^{\pm}$, and the current $J^{\mathcal{B}_{2}^{*}}$ can couple both to baryons with $J^{P}=\frac{3}{2}^{\pm}$ and $J^{P}=\frac{1}{2}^{\pm}$. To extract the form factors of the decay process $\frac{1}{2}^{+}\rightarrow\frac{3}{2}^{+}$, we must eliminate the interferences of the baryons with spin-parities $\frac{1}{2}^{\pm}$, $\frac{3}{2}^{-}$ in final states, and $\frac{1}{2}^{-}$ in initial states. It is shown by Eq. (\ref{eq:5}) that the current $J^{\mathcal{B}_{2}^{*}}$ coupling to baryons with $J^{P}=\frac{1}{2}^{\pm}$ are related to the structures of $\gamma_{\mu}$ and $p_{\mu}^{\prime}$. Hence, the structures proportional to $\gamma_{\mu}$ and $p_{\mu}^{\prime}$ must be eliminated firstly. After these interferences are eliminated, the correlation function can be written as Eq. (\ref{eq:A2}) in Appendix \ref{Sec:AppA}, and can be decomposed into following different structures,
\begin{eqnarray}\label{eq:8}
\Pi_{\mu\nu}^{\mathrm{phy-V/A}}=\sum^{16}_{1}\Pi_{i}^{\mathrm{phy-V/A}}e_{i\mu\nu}
\end{eqnarray}
where
\begin{eqnarray}\label{eq:9}
\notag
&&e_{i\mu\nu}=\gamma_{\nu}\times\{1,\slashed p,\slashed p^{\prime}\}\times\gamma_{5}p_{\mu} \quad i=1\sim4 \\ \notag
&&e_{i\mu\nu}= \{1,\slashed p,\slashed p^{\prime}\}\times\gamma_{5}g_{\mu\nu} \quad i=5\sim8 \\ \notag
&&e_{i\mu\nu}= \{1,\slashed p,\slashed p^{\prime}\}\times\gamma_{5}p_{\mu}p_{\nu} \quad i=9\sim12 \\
&&e_{i\mu\nu}= \{1,\slashed p,\slashed p^{\prime}\}\times\gamma_{5}p_{\mu}p^{\prime}_{\nu} \quad i=13\sim16
\end{eqnarray}
are corresponding to vector part $\Pi_{i}^{\mathrm{phy-V}}$ and
\begin{eqnarray}\label{eq:10}
\notag
&&e_{i\mu\nu}=\gamma_{\nu}\times\{1,\slashed p,\slashed p^{\prime}\}\times p_{\mu} \quad i=1\sim4 \\ \notag
&&e_{i\mu\nu}= \{1,\slashed p,\slashed p^{\prime}\}\times g_{\mu\nu} \quad i=5\sim8 \\ \notag
&&e_{i\mu\nu}= \{1,\slashed p,\slashed p^{\prime}\}\times p_{\mu}p_{\nu} \quad i=9\sim12 \\
&&e_{i\mu\nu}= \{1,\slashed p,\slashed p^{\prime}\}\times p_{\mu}p^{\prime}_{\nu} \quad i=13\sim16
\end{eqnarray}
are matched with the axial vector part $\Pi_{i}^{\mathrm{phy-A}}$.
The form factors are included in these expansion coefficients $\Pi^{\mathrm{phy-V/A}}_i$ which are commonly named as scalar invariant amplitudes. In order to eliminate the contributions of baryons with negative parity, all these 32 structures will be used to determine the value of form factors for the process $\frac{1}{2}^{+}\rightarrow\frac{3}{2}^{+}$.

In QCD side, we substitute the interpolating currents in the three-point correlation function in Eq. (\ref{eq:2}) with their specific form in Eq. (\ref{eq:3}). Then, the correlation function in QCD side can be expressed as the following form after doing the operator production expansion (OPE),
\begin{eqnarray}\label{eq:11}
\notag
&&\Pi_{\mu\nu} ^{\mathrm{QCD-V/A}}(p,p')= 2\sqrt{2}i^{2}\varepsilon _{nml}\varepsilon _{abc}\int d^4xd^4ye^{ip'x}e^{iqy}\Big\{ Q^{la}(x) \\
\notag
&& \times \gamma _{\alpha}\mathcal{C}Q^{i'bT}(y)\mathcal{C}\gamma _\nu(1-\gamma_{5})\mathcal{C}q^{\prime miT}(x - y)\mathcal{C}\gamma _{\mu}q^{nc}(x)\gamma_{5}\gamma_{\alpha}\Big\} \\
\end{eqnarray}
where $Q^{ij}(x)$ is the full propagator of $c$ quark and can be written as the following form in momentum space~\cite{Pascual,Reinders},
\begin{eqnarray}\label{eq:12}
\notag
Q^{ij}(x) &&= \frac{i}{(2\pi )^4}\int d^4k e^{- ikx}\left\{ \frac{\delta ^{ij}}{\slashed k - m_{Q}} \right.\\
\notag
&& - \frac{{{g_s}G_{\alpha \beta }^nt_{ab}^n}}{4}\frac{{{\sigma ^{\alpha \beta }}(\slashed{k} + {m_{Q}}) + (\slashed{k} + {m_{Q}}){\sigma ^{\alpha \beta }}}}{{{{({k^2} - m_{Q}^2)}^2}}}\\
\notag
&& - \frac{{g_s^2{{({t^m}{t^n})}_{ab}}G_{\alpha \beta }^mG_{\mu \nu }^n({f^{\alpha \beta \mu \nu }} + {f^{\alpha \mu \beta \nu }} + {f^{\alpha \mu \nu \beta }})}}{{4{{({k^2} - m_{Q}^2)}^5}}} \\
&&\left. { + \cdots} \right\}
\end{eqnarray}
with
\begin{eqnarray}
\notag
{f^{\alpha \beta \mu \nu }} &&= (\slashed k + {m_{Q}}){\gamma ^\alpha }(\slashed k + {m_{Q}}){\gamma ^\beta }(\slashed k + {m_{Q}})\\ \notag
&& \times {\gamma ^\mu }(\slashed k + {m_{Q}}){\gamma ^\nu }(\slashed k + {m_{Q}})
\end{eqnarray}
In Eq. (\ref{eq:11}), $q^{ij}(x)$ and $q^{\prime ij}(x-y)$ are light quark full propagators of $u$, $d$ or $s$ and they can be uniformly expressed as the following form in coordinate space~\cite{Pascual,Reinders},
\begin{eqnarray}\label{eq:13}
\notag
q^{ij}(x) && = \frac{i\delta ^{ij}\slashed{x}}{2\pi ^2x^4} - \frac{\delta ^{ij}\left\langle \bar qq \right\rangle }{12} - \frac{\delta ^{ij}x^2\left\langle \bar qg_s\sigma Gq \right\rangle }{192}\\
\notag
&& - \frac{\left\langle {\bar q^j\sigma ^{\mu \nu }q^i} \right\rangle \sigma _{\mu \nu }}{8} - \frac{ig_sG_{\alpha \beta }^nt_{ij}^n(\slashed{x}\sigma ^{\alpha \beta } + \sigma ^{\alpha \beta}\slashed x)}{32\pi ^2x^2}\\
&& - \frac{i\delta ^{ij}x^2\slashed{x}g_s^2\left\langle \bar qq \right\rangle ^2}{7776} +\cdots
\end{eqnarray}
Here, $\left\langle \bar qg_s\sigma Gq \right\rangle=\left\langle \bar qg_s\sigma_{\mu\nu}t^\alpha G^\alpha_{\mu\nu}q \right\rangle$, $t^{\alpha}=\frac{\lambda^\alpha}{2}$, $\lambda^{\alpha}$ ($\alpha=1,...,8$) are the Gell-Mann matrices, and $\sigma_{\mu\nu}=\frac{i}{2}[\gamma_\mu,\gamma_\nu]$. By using the double dispersion relation, the correlation function in QCD side can be written as,
\begin{eqnarray}\label{eq:14}
\Pi _{\mu\nu}^{\mathrm{QCD-V/A}}(p,p') &&= \int\limits_{s_{\min}}^\infty  {ds} \int\limits_{u_{\min}}^\infty  {du} \frac{{\rho _{\mu\nu} ^{\mathrm{QCD-V/A}}(s,u,q^2)}}{{(s - p^2)(u - p'^2)}}
\end{eqnarray}
where $\rho^{\mathrm{QCD}}_{\mu\nu}(s,u,q^2)$ is QCD spectral density with $s=p^2$ and $u=p'^2$. $s_{\min}$ and $u_{\min}$ are create thresholds for initial and final baryons which are taken as $4m_{c}^{2}$ and $m_{c}^{2}$, respectively.
Eliminating the structures proportional to $\gamma_{\mu}$ and $p_{\mu}^{\prime}$, the correlation function in QCD side can also be decomposed into the same structures as those in phenomenological side,
\begin{eqnarray}\label{eq:15}
\Pi_{\mu\nu}^{\mathrm{QCD-V/A}}=\sum^{16}_{1}\Pi_{i}^{\mathrm{QCD-V/A}}e_{i\mu\nu}
\end{eqnarray}
\begin{figure*}[htbp]
{\includegraphics[width=1\textwidth]{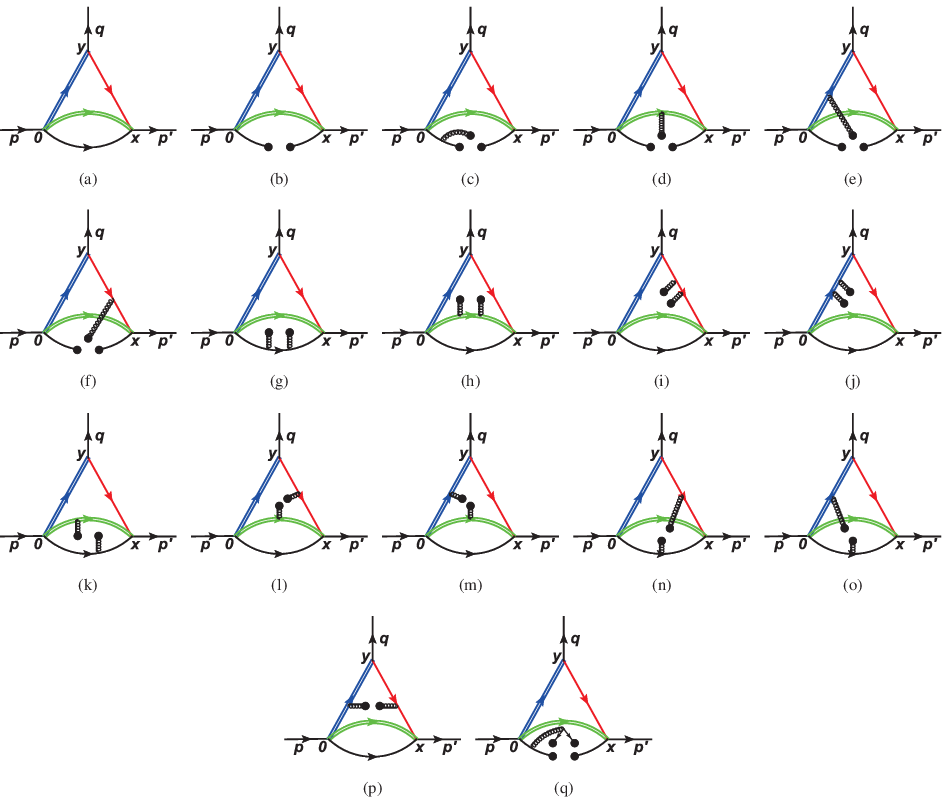}}
\caption{ The Feynman diagrams for the perturbative part and vacuum
condensate terms, where the doubly-solid line denotes a charm quark, and the ordinary solid line represents a light quark.}
\label{feynman}
\end{figure*}
where $e_{i\mu\nu}$ has the same form as Eqs. (\ref{eq:9}) and (\ref{eq:10}) for $\Pi_{\mu\nu}^{\mathrm{QCD-V}}$ and $\Pi_{\mu\nu}^{\mathrm{QCD-A}}$, respectively.
In Eq. (\ref{eq:15}), $\Pi^{\mathrm{QCD-V}}_i$ and $\Pi^{\mathrm{QCD-A}}_i$ ($i=1,2\cdots16$) are scalar invariant amplitudes in QCD side. These scalar invariant amplitudes can be represented as the summation of perturbative part and vacuum condensate terms. The Feynman diagrams about the perturbative part, vacuum condensate terms of $\langle{\bar qq}\rangle$, $\langle g_{s}^{2}GG\rangle$, $\langle \bar q g_{s}\sigma Gq\rangle$ and $g_{s}^{2}\langle{\bar qq}\rangle^{2}$ are explicitly shown in Fig. \ref{feynman}.

\subsection{The QCD spectral density in the QCD side}\label{sec2.3}

For the correlation function which is related to the vector transition current $\overline{q}^{\prime}\gamma_{\nu}c$ as an example, we now introduce how its spectral density $\rho^{\mathrm{QCD-V}}_{\mu\nu}$ in Eq. (\ref{eq:14}) is calculated. Firstly, the contribution of perturbative term is derived by substituting the free propagators in momentum space of both light and heavy quarks in Eq. (\ref{eq:11}). The corresponding Feynman diagram is shown as Fig. \ref{feynman} (a). After performing integrals on coordinates $x$ and $y$, we write the perturbative part as,
\begin{eqnarray}\label{eq:16}
\notag
&&\Pi _{\mu\nu}^{\mathrm{QCD-V0}}(p,p') =- \frac{12\sqrt{2}}{{{{(2\pi )}^8}}}\int {{d^4}{k_1}} {d^4}{k_2}{d^4}{k_3}{d^4}{k_4}\int d^{4}q^{\prime}\\
&& \times\frac{{\delta ^4}(q^{\prime} - {k_1} - {k_4}) {\delta ^4}(p' - q^{\prime} - {k_3}){\delta ^4}(q - {k_2}+k_{3})}{{(k_{1}^{2}-m_{1}^{2})(k_{2}^{2}-m_{2}^{2})(k_{3}^{2}-m_{3}^{2})(k_{4}^{2} - m_{4}^{2})}}N_{\mu\nu}
\end{eqnarray}
where $N_{\mu\nu}=({{\slashed k}_1}+m_{1})\gamma _{\alpha}({{\slashed k}_2}-m_{2}){\gamma _\nu }({{\slashed k}_3}-m_{3}){\gamma _\mu}({{\slashed k}_4} + {m_{4}}){\gamma _5}\gamma _{\alpha}$. The superscript $\mathrm{0}$ denotes that the dimension of perturbative term is zero. By setting all quark lines on-shell with the Cutkosky's rule~\cite{Cutkosky}, the QCD spectral density function of this contribution can be derived as,
\begin{eqnarray}\label{eq:17}
\notag
&&\rho _{\mu\nu} ^{\mathrm{QCD-V0}}(s,u,{q^2})=-\frac{{12\sqrt{2}}}{{{{(2\pi )}^8}}}\frac{{{{( - 2\pi i)}^5}}}{{{{(2\pi i)}^3}}}\\ \notag
&& \times\int\limits_{(m_{1}+m_{4})^2}^{(\sqrt{u}-m_{3})^2} {dr^{\prime}} \int {{d^4}{k_1}}\delta (k_1^2-m_{1}^{2})\delta [{(q' - {k_1})^2}-m_{4}^{2}]  \\ \notag
&&\times\int {{d^4}{k_3}} \delta [{(p' - {k_3})^2} - r^{\prime} ]\delta (k_3^2-m_{3}^{2})\delta [{({k_3} + p-p^{\prime})^2} - m_{2}^2]\\
&& \times N_{\mu\nu}
\end{eqnarray}
where $q'=k_{1}+k_{4}$ and $r'=q'^{2}$. The integral formulas about two and three Dirac delta functions in this equation are shown in Appendix \ref{Sec:AppB}.

The quark condensate $\langle\bar qq\rangle$ has dimension 3 and it comes from the second term of the light quark full propagator in Eq. (\ref{eq:13}). Because heavy quark do not contribute to this condensation, there is only one Feynman diagram originating from light quark condensate in the present work. The corresponding Feynman diagram is shown in Fig. \ref{feynman} (b) which can be expressed as,
\begin{eqnarray}\label{eq:18}
\notag
&&\Pi _{\mu\nu }^{\mathrm{QCD-V3}}(p,p')= \frac{i\sqrt{2}\langle \overline{q}q\rangle}{(2\pi)^4}\times\\ \notag
&&\int d^{4}k_{3}\frac{1}{\left[\left(p^{\prime}-k_{3}\right)^{2}-m_{1}^{2}\right]
\left[\left(p-p^{\prime}+k_{3}\right)^{2}-m_{2}^{2}\right]\left[k_{3}^{2}-m_{3}^{2}\right]}\times\\ \notag
&& \Big\{\left[\slashed p^{\prime}-\slashed k_{3}+m_{1}\right]\gamma_{\alpha}\left[\slashed p-\slashed p^{\prime}+\slashed k_{3}-m_{2}\right]\gamma_{\nu}\left[\slashed k_{3}-m_{3}\right]\gamma_{\mu}\gamma_{5}\gamma_{\alpha}\Big\} \\
\end{eqnarray}
According to the Cutkosky's rule, its QCD spectral density can be simplified into the integration of three delta functions,
\begin{eqnarray}\label{eq:19}
\notag
&&\rho _{\mu\nu }^{\mathrm{QCD-V3}}(s,u,q^{2})= \frac{i\sqrt{2}\langle \overline{q}q\rangle}{(2\pi)^4}\frac{\left(-2\pi i\right)^{3}}{\left(2\pi i\right)^{2}}\int d^{4}k_{3}\delta\left[k_{3}^{2}-m_{3}^{2}\right]\\
&&\times\delta\left[\left(p^{\prime}-k_{3}\right)^{2}-m_{1}^{2}\right]
\delta\left[\left(p-p^{\prime}+k_{3}\right)^{2}-m_{2}^{2}\right]N_{\mu\nu}^{V3}
\end{eqnarray}
where
\begin{eqnarray}
\notag
N_{\mu\nu}^{V3}&&=\left(\slashed p^{\prime}-\slashed k_{3}+m_{1}\right)\gamma_{\alpha}\left(\slashed p-\slashed p^{\prime}+\slashed k_{3}-m_{2}\right)\gamma_{\nu}\left(\slashed k_{3}-m_{3}\right)\\ \notag
&&\times\gamma_{\mu}\gamma_{5}\gamma_{\alpha}
\end{eqnarray}

For quark gluon mixed condensate $\langle\bar{q}g_{s}\sigma Gq\rangle$, there are two mechanisms to produce this term. These two kinds of mixed condensates are illustrated in Fig. \ref{feynman} (c) and (d)$\sim$(f), respectively. The contribution of Fig. \ref{feynman} (c) originates from the third term of light quark full propagator in Eq. (\ref{eq:13}). To finish the integration of the correlation function related to this contribution (Fig. \ref{feynman} (c)) in coordinate space, the coordinate $x_{\mu}$ is firstly replaced by $x_{\mu}\to i\partial /\partial p_{\mu}^{\prime}$. Then, after performing the derivative on $p^{\prime}_{\mu}$, there will appear a high power factor $\frac{1}{(k^{2}-m^{2})^{n}}$ in Feynman integral. To obtain the spectral density with the Cutkosky's rule, we adopt the following formula to lower the power of the denominator,
\begin{eqnarray}\label{eq:20}
\frac{1}{(k^{2}-m^{2})^{n}}=\frac{1}{(n-1)!}\frac{\partial^{n-1}}{(\partial d)^{n-1}}\frac{1}{k^{2}-d}\Bigg|_{d\rightarrow m^{2}}
\end{eqnarray}
With Eq. (\ref{eq:20}), this kind of quark gluon mixed condensate can be written as,
\begin{eqnarray}\label{eq:21}
\notag
&&\Pi _{\mu\nu }^{\mathrm{QCD-V5}}(p,p')=-\frac{i\sqrt{2}\langle \overline{q}g_{s}\sigma Gq\rangle}{16(2\pi)^4}\Bigg\{\frac{\partial}{\partial d}\int d^{4}k_{3}\times\\ \notag
&&\frac{N^{V5}_{\mu\nu1}}{\left[\left(p^{\prime}-k_{3}\right)^{2}-d\right]
\left[\left(p-p^{\prime}+k_{3}\right)^{2}-m_{2}^{2}\right]\left[k_{3}^{2}-m_{3}^{2}\right]}\\ \notag
&&+ \frac{\partial^{2}}{2\partial d^{2}}\int d^{4}k_{3}\frac{N^{V5}_{\mu\nu2}}{\left[\left(p^{\prime}-k_{3}\right)^{2}-d\right]\left[\left(p-p^{\prime}+k_{3}\right)^{2}-m_{2}^{2}\right]
}\\
&&\times\frac{1}{\left[k_{3}^{2}-m_{3}^{2}\right]}\Bigg\}_{d\rightarrow m_{1}^{2}}
\end{eqnarray}
The second kind of mixed condensate originates from two quark lines, which can be seen in Figs. \ref{feynman} (d)$\sim$(f). It is shown that one gluon emitted from a light or heavy quark line together with a quark-antiquark pair contributed from another light quark line form into this condensation. This kind of contribution comes from the second term in Eq. (\ref{eq:12}) and the fourth term in Eq. (\ref{eq:13}). For Fig. \ref{feynman} (d) as an example, it can finally be expressed as,
\begin{eqnarray}\label{eq:22}
\notag
&&\Pi _{\mu\nu }^{\mathrm{QCD-V5}}(p,p')=\frac{i\sqrt{2}\langle \overline{q}g_{s}\sigma Gq\rangle}{192\left(2\pi\right)^{4}}\frac{\partial}{\partial d}\Bigg\{\int d^{4}k_{3}\times \\ \notag
&&\frac{N^{V5}_{\mu\nu3}}{\left[\left(p^{\prime}-k_{3}\right)^{2}-d\right]
\left[\left(p-p^{\prime}+k_{3}\right)^{2}-m_{2}^{2}\right]\left[k_{3}^{2}-m_{3}^{2}\right]}\Bigg\}_{d\rightarrow m_{1}^{2}} \\
\end{eqnarray}
Similar to quark condensate, the QCD spectral density of these mixed condensate terms can also be derived as,
\begin{eqnarray}\label{eq:23}
\notag
&&\rho _{\mu\nu }^{\mathrm{QCD-V5}}(s,u,q^{2})= \frac{i\sqrt{2}\langle \overline{q}g_{s}\sigma Gq\rangle}{16(2\pi)^4}\frac{\left(-2\pi i\right)^{3}}{\left(2\pi i\right)^{2}}\times\\ \notag
&&\Bigg\{\frac{\partial}{\partial d}\int d^{4}k_{3}\delta\left[\left(p^{\prime}-k_{3}\right)^{2}-d\right]
\delta\left[\left(p-p^{\prime}+k_{3}\right)^{2}-m_{2}^{2}\right]\\ \notag
&&\times \delta\left[k_{3}^{2}-m_{3}^{2}\right]N^{V5}_{\mu\nu1}+\frac{\partial^{2}}{2\partial d^{2}}\int d^{4}k_{3}\delta\left[\left(p^{\prime}-k_{3}\right)^{2}-d\right] \\
&&
\delta\left[\left(p-p^{\prime}+k_{3}\right)^{2}-m_{2}^{2}\right]\delta\left[k_{3}^{2}-m_{3}^{2}\right] N^{V5}_{\mu\nu2}\Bigg\}_{d\rightarrow m_{1}^{2}}
\end{eqnarray}
and
\begin{eqnarray}\label{eq:24}
\notag
&&\rho _{\mu\nu }^{\mathrm{QCD-V5}}(s,u,q^{2})=\\ \notag
&&\frac{i\sqrt{2}\langle \overline{q}g_{s}\sigma Gq\rangle}{192\left(2\pi\right)^{4}}\frac{\left(-2\pi i\right)^{3}}{\left(2\pi i\right)^{2}}\frac{\partial}{\partial d}\Bigg\{\int d^{4}k_{3}N^{V5}_{\mu\nu3}\times \\ \notag
&&\delta\left[\left(p^{\prime}-k_{3}\right)^{2}-d\right]
\delta\left[\left(p-p^{\prime}+k_{3}\right)^{2}-m_{2}^{2}\right]\delta\left[k_{3}^{2}-m_{3}^{2}\right]\Bigg\}_{d\rightarrow m_{1}^{2}} \\
\end{eqnarray}
with
\begin{eqnarray}
\notag
N^{V5}_{\mu\nu1}&&=-4\left[\slashed p^{\prime}-\slashed k_{3}\right]\gamma_{\alpha}\left[\slashed p-\slashed p^{\prime}+\slashed k_{3}-m_{2}\right]\gamma_{\nu}\left[\slashed k_{3}-m_{3}\right]\\ \notag
&&\times\gamma_{\mu}\gamma_{5}\gamma_{\alpha} \\ \notag
N^{V5}_{\mu\nu2}&&=8\left[\left(\slashed p^{\prime}-\slashed k_{3}+m_{1}\right)m_{1}^{2}\right]\gamma_{\alpha}\left[\slashed p-\slashed p^{\prime}+\slashed k_{3}-m_{2}\right]\gamma_{\nu}\\ \notag
&&\times\left[\slashed k_{3}-m_{3}\right]\gamma_{\mu}\gamma_{5}\gamma_{\alpha}
\end{eqnarray}
\begin{eqnarray}
\notag
&&N^{V5}_{\mu\nu3}=\\ \notag
&&\left(g_{\lambda\beta}g_{\rho\tau}-g_{\lambda\tau}g_{\rho\beta}\right)\left[\sigma^{\beta\tau}\left(\slashed p^{\prime}-\slashed k_{3}+m_{1}\right)+\left(\slashed p^{\prime}-\slashed k_{3}+m_{1}\right)\sigma^{\beta\tau}\right]\\ \notag
&&\times\gamma_{\alpha}\left[\slashed p-\slashed p^{\prime}+\slashed k_{3}-m_{2}\right]\gamma_{\nu}\left(k_{3}-m_{3}\right)\gamma_{\mu}\sigma^{\lambda\rho}\gamma_{5}\gamma_{\alpha}
\end{eqnarray}

Based on the generation mechanism of the gluon condensate $\langle g_{s}^{2}GG\rangle$, they can also be divided into two categories which are are shown in Figs. \ref{feynman} (g) $\sim$ (j) and (k) $\sim$ (p), respectively. For the contributions of Fig. \ref{feynman} (g) and (m) as examples, they finally can be expressed as,
\begin{eqnarray}\label{eq:25}
\notag
&&\Pi _{\mu\nu }^{\mathrm{QCD-V4}}(p,p')=\frac{\sqrt{2}\langle g_{s}^{2}GG\rangle}{24\left(2\pi\right)^{8}}\int\limits^{\left(\sqrt{u}-m_{3}^{2}\right)^{2}}_{\left(m_{1}+m_{4}\right)^{2}}dr^{\prime}\times \\ \notag
&&\Bigg\{\int d^{4}k_{3}\frac{1}{\left[k_{3}^{2}-m_{3}^{2}\right]\left[\left(p^{\prime}-k_{3}\right)^{2}-r^{\prime}\right]
\left[\left(p-p^{\prime}+k_{3}\right)^{2}-m_{2}^{2}\right]}\\
&&\times\frac{\partial^{4}}{4!\partial d^{4}}\int d^{4}k_{1}\frac{N^{V4}_{\mu\nu1}}{\left[k_{1}^{2}-d\right]\left[\left(q^{\prime}-k_{1}\right)^{2}-m_{4}^{2}\right]}\Bigg\}_{d\rightarrow m_{1}^{2}}
\end{eqnarray}
and
\begin{eqnarray}\label{eq:26}
\notag
&&\Pi _{\mu\nu }^{\mathrm{QCD-V4}}(p,p')=\frac{\sqrt{2}\langle g_{s}^{2}GG\rangle}{192\left(2\pi\right)^{8}}\int\limits^{\left(\sqrt{u}-m_{3}^{2}\right)^{2}}_{\left(m_{1}+m_{4}\right)^{2}}dr^{\prime}\times \\ \notag
&&\Bigg\{\frac{\partial}{\partial B}\int d^{4}k_{3}\frac{1}{\left[k_{3}^{2}-m_{3}^{2}\right]\left[\left(p^{\prime}-k_{3}\right)^{2}-r^{\prime}\right]
\left[\left(p-p^{\prime}+k_{3}\right)^{2}-B\right]}\\
&&\times\frac{\partial}{\partial A}\int d^{4}k_{1}\frac{N^{V4}_{\mu\nu2}}{\left[k_{1}^{2}-A\right]
\left[\left(q^{\prime}-k_{1}\right)^{2}-m_{4}^{2}\right]}\Bigg\}_{A\rightarrow m_{1}^{2},B\rightarrow m_{2}^{2}}
\end{eqnarray}
with
\begin{eqnarray}
\notag
N^{V4}_{\mu\nu1}&&=\left(g_{\lambda\beta}g_{\rho\tau}-g_{\lambda\tau}g_{\rho\beta}\right)\left(f^{\lambda\rho\beta\tau}+f^{\lambda\beta\rho\tau}+f^{\lambda\beta\tau\rho}\right)\gamma_{\alpha}\\ \notag &&\times\left[\slashed p-\slashed p^{\prime}+\slashed k_{3}-m_{2}\right]\gamma_{\nu}\left[\slashed k_{3}-m_{3}\right]\gamma_{\mu}\gamma_{5}\gamma_{\alpha}
\end{eqnarray}
\begin{eqnarray}
\notag
&&N_{\mu\nu2}^{V4}=\left(g_{\lambda\beta}g_{\rho\tau}-g_{\lambda\tau}g_{\rho\beta}\right)\left[\sigma^{\lambda\rho}\left(\slashed k_{1}+m_{1}\right)+\left(\slashed k_{1}+m_{1}\right)\sigma^{\lambda\rho}\right]
\\ \notag
&&\times\gamma_{\alpha}\left[\left(\slashed p-\slashed p^{\prime}+\slashed k_{3}-m_{2}\right)\sigma^{\tau\beta}+\sigma^{\tau\beta}\left(\slashed p-\slashed p^{\prime}+\slashed k_{3}-m_{2}\right)\right]\\ \notag
&&\times\gamma_{\nu}\left[\slashed k_{3}-m_{3}\right]\gamma_{\mu}\left(\slashed p^{\prime}-k_{3}-k_{1}+m_{4}\right)\gamma_{5}\gamma_{\alpha}
\end{eqnarray}
The spectral density of these above gluon condensates can also be derived by the Cutkosky's rule. Their final expressions are similar to that of pertubertive term (Eq. (\ref{eq:17})). The finally considered condensate term in the present work is the four quark condensation which is shown in Fig. \ref{feynman} (q). This figure describes that a light quark line emits a quark-antiquark pair and a gluon, and then the gluon contributes a quark-antiquark pair to produce the four quark condensation. This contribution comes from the sixth term in Eq. (\ref{eq:13}) and can be calculated by doing variable replacement $x_{\mu}\to i\partial /\partial p{'_\mu }$. This contribution can be expressed as the following form,

\begin{eqnarray}\label{eq:27}
\notag
&&\Pi _{\mu\nu }^{\mathrm{QCD-V6}}(p,p')=\frac{i\sqrt{2} g_{s}^{2}\langle \overline{q}q\rangle^{2}}{648\left(2\pi\right)^{4}}
\Bigg\{\frac{\partial}{\partial d}\int \frac{N^{V6}_{\mu\nu1}}{M(d,k_{3})}d^{4}k_{3} \\ 
&&+\frac{\partial^{2}}{2\partial d^{2}}\int \frac{N^{V6}_{\mu\nu2}}{M(d,k_{3})}d^{4}k_{3}+\frac{\partial^{3}}{6\partial d^{3}}\int \frac{N^{V6}_{\mu\nu3}}{M(d,k_{3})}d^{4}k_{3}\Bigg\}_{d\rightarrow m_{1}^{2}}
\end{eqnarray}

with
\begin{eqnarray}
\notag
M(d,k_{3})=\left[k_{3}^{2}-m_{3}^{2}\right]\left[\left(p^{\prime}-k_{3}\right)^{2}-d\right]
\left[\left(p-p^{\prime}+k_{3}\right)^{2}-m_{2}^{2}\right]
\end{eqnarray}
\begin{eqnarray}
\notag
N^{V6}_{\mu\nu1}=-12 \gamma^{\beta }{\gamma }^{\alpha }\left(\slashed{p}-\slashed{p}^{\prime}+\slashed{k}_{3}-m_2\right){\gamma }_{\nu }\left(\slashed{k}_{3}-m_{3}\right){\gamma }_{\mu }{\gamma }_{\beta }{\gamma }^{5}{\gamma }_{\alpha }
\end{eqnarray}
\begin{eqnarray}
\notag
N^{V6}_{\mu\nu2}&&=\Bigl(64 \slashed{k}_{3} {k}_{3}^{\beta } - 48 m_{1}{k}_{3}^{\beta }-16\left({k}_{3}\cdot{p^{\prime}}\right){\gamma }^{\beta }
-64\slashed{k}_{3}{p^{\prime}}^{\beta } \\ \notag
&&-64 \slashed{p}' {k}_3^{\beta }+8m_{3}^{2}{\gamma }^{\beta }+ 48 m_{1}{p^{\prime}}^{\beta } + 8 {p^{\prime}}^2{\gamma }^{\beta } + 64 \slashed p^{\prime} {p^{\prime}}^{\beta }\Bigr) \\ \notag
&&\times\left(\slashed p^{\prime}-\slashed{k}_3+m_1\right){\gamma }^{\alpha }\left(\slashed{p}-\slashed p^{\prime}+\slashed{k}_{3}-m_2\right){\gamma }_{\nu }\left(\slashed{k}_{3}-m_{3}\right)\\ \notag
&&\times{\gamma }_{\mu }{\gamma }_{\beta }{\gamma }_5{\gamma }_{\alpha }
\end{eqnarray}
\begin{eqnarray}
\notag
N^{V6}_{\mu\nu3}&&=\Bigl(48 m_3^2  \left(m_1-\slashed{k}_3\right){k}_3^{\beta } + 96  \left(m_1-\slashed{k}_3\right) \left(  {k}_3\cdot {p'}\right){p'}^{\beta } \\ \notag
&&- 96  \left(m_1-\slashed{k}_3\right) \left(  {k}_3\cdot {p'}\right){k}_3^{\beta } - 48 m_3^2  \left(m_1-\slashed{k}_3\right){p'}^{\beta } \\ \notag
&&+ 48  {p'}^2 \left(m_1-\slashed{k}_3\right){k}_3^{\beta } - 48 {p'}^2  \left(m_1-\slashed{k}_3\right){p'}^{\beta }+ 48 m_3^2  \slashed{p}'{k}_3^{\beta } \\ \notag
&& + 96  \left( {k}_3 \cdot {p'}\right) \slashed{p}' {p'}^{\beta }- 96  \left(  {k}_3 \cdot{p'}\right) \slashed{p}'{k}_3^{\beta } + 48  {p'}^2 \slashed{p}' {k}_3^{\beta } \\ \notag
&& - 48 m_3^2  \slashed{p}'{p'}^{\beta }- 48 {p'}^2 \slashed{p}'{p'}^{\beta }\Bigr)\left(\slashed{p}'-\slashed{k}_3 + m_1\right){\gamma }^{\alpha }  \\ \notag
&&\times\left(\slashed{p}-\slashed{p}'+\slashed{k}_3-m_2\right){\gamma }_{\nu }\left(\slashed{k}_3 - m_3\right){\gamma }_{\mu }{\gamma }_{\beta }{\gamma }_5{\gamma }_{\alpha }
\end{eqnarray}
Its spectral density function can also be obtained by Cutkosky's rule and has similar form as that of quark condensate term.

Finally, we take the change of variables $p^2\to-P^2$, $p'^2\to-P'^2$ and $q^2\to-Q^2$ in Eqs. \ref{eq:A2} and \ref{eq:14}, and perform double Borel transformation for variables $P^2$ and $P'^2$ to both phenomenological and QCD sides. The variables $P^2$ and $P'^2$ are then replaced by Borel parameters $\mathrm{M}_1^2$ and $\mathrm{M}_2^2$. It is shown by Eqs. (\ref{eq:8}) and (\ref{eq:15}) that there are both 16 dirac structures at the phenomenological and QCD sides. Using quark-hadron duality condition, we can establish 16 linear equations about form factors. By solving these linear equations, we obtain the momentum dependent form factors $F_{i}(Q^2)$ and $G_{i}(Q^2)$ which are shown in Eq. (\ref{eq:A3}) in Appendix \ref{Sec:AppA}. In Eq. (\ref{eq:A3}), $s_{0}$ and $u_{0}$ are threshold parameters for initial and final state baryons, respectively. The spectral density functions $\rho_{i}^{\mathrm{QCD-V/A}}(s,u,Q^2)$ include contributions of perturbative part and different condensate terms, and $i=1\sim16$ represent 16 different dirac structures in QCD side. The full expressions of these functions are too complex to be shown here for simplicity.

\section{Numerical results}\label{sec3}
From Eq. (\ref{eq:A3}), one can find that the final results of QCDSR depend on some parameters such as the Borel parameters $\mathrm{M}_{1}^{2}$, $\mathrm{M}_{2}^{2}$, the threshold parameters $s_0$, $u_0$, the vacuum condensate terms, the pole residues and the masses of quark and hadrons. Most of the input parameters in the present work are all listed in Tab. \ref{PV}.
\begin{table*}[htbp]
\caption{1 Input parameters (IP) in this work. The values of vacuum condensate are at the energy scale $\mu=1$ GeV.}
\begin{ruledtabular}
\renewcommand{\arraystretch}{1.3}
\label{PV}
\begin{tabular}{c| c c c c c}
 IP&$m_{\Xi_{cc}(\frac{1}{2}^{+})}$&$m_{\Xi_{cc}(\frac{1}{2}^{-})}$& $m_{\Omega_{cc}(\frac{1}{2}^{+})}$&$m_{\Omega_{cc}(\frac{1}{2}^{-})}$ & $m_{\Sigma_{c}^{*}(\frac{3}{2}^{+})}$\\
Values(GeV)& $3.621$ \cite{Yu:2022lel}&$3.932$ \cite{Yu:2022lel}&$3.750$ \cite{Yu:2022lel}&$4.049$ \cite{Yu:2022lel}& $2.518$ \cite{Yu:2022ymb}\\ \hline
IP&$m_{\Sigma_{c}^{*}(\frac{3}{2}^{-})}$&$m_{\Xi_{c}^{*\prime}(\frac{3}{2}^{+})}$& $m_{\Xi_{c}^{*\prime}(\frac{3}{2}^{-})}$&$m_{\Omega_{c}^{*}(\frac{3}{2}^{+})}$ & $m_{\Omega_{c}^{*}(\frac{3}{2}^{-})}$\\
Values(GeV)& $2.829$ \cite{Yu:2022ymb}&$2.645$ \cite{Yu:2022ymb}& $2.958$ \cite{Yu:2022ymb}& $2.762$ \cite{Li:2022xtj}&$3.062$ \cite{Li:2022xtj}\\ \hline
IP&$\lambda_{\Xi_{cc}(\frac{1}{2}^{+})}$&$\lambda_{\Omega_{cc}(\frac{1}{2}^{+})}$ & $\lambda_{\Sigma_{c}^{*}(\frac{3}{2}^{+})}$& $\lambda_{\Xi_{c}^{*\prime}(\frac{3}{2}^{+})}$&	$\lambda_{\Omega_{c}^{*}(\frac{3}{2}^{+})}$	\\
Values(GeV$^{3}$)& $0.109\pm0.020$ \cite{Shi:2019hbf}&$0.129\pm0.024$ \cite{Shi:2019hbf}&$\sqrt{2}\times(0.027\pm0.008)$ \cite{Wang:2010vn}&$\sqrt{2}\times(0.033\pm0.008)$ \cite{Wang:2010vn}&$0.056\pm0.012$	\cite{Wang:2010vn}\\ \hline
IP& $m_{e}$ & $m_{\mu}$ & $\langle \overline{q}q\rangle$ & $\langle \overline{s}s\rangle$  & $\langle \overline{q}g_{s}\sigma Gq\rangle$  \\
Values& $0.511\times10^{-3}$ GeV \cite{ParticleDataGroup:2024cfk}&$106\times10^{-3}$ GeV \cite{ParticleDataGroup:2024cfk}& $-(0.23\pm0.01)^{3}$ GeV$^{3}$ \cite{Shifman:1978by,Reinders}& $(0.8\pm0.1)\langle \overline{q}q\rangle$ \cite{Shifman:1978by,Reinders}&$m_{0}^{2}\langle \overline{q}q\rangle$ \cite{Shifman:1978by,Reinders}\\ \hline
IP&$\langle \overline{s}g_{s}\sigma Gs\rangle$ & $m_{0}^{2}$ & $\langle g_{s}^{2}GG\rangle$ &  \\
Values&$m_{0}^{2}\langle \overline{s}s\rangle$ \cite{Shifman:1978by,Reinders}& $0.8\pm0.1$ GeV$^{2}$ \cite{Shifman:1978by,Reinders}&$0.47\pm0.15$ GeV$^{4}$ \cite{Narison:2010cg,Narison:2011xe,Narison:2011rn}& & \\
\end{tabular}
\end{ruledtabular}
\end{table*}
We know that the mass of heavy quark and the values of vacuum condensates are energy dependent according to the renormlization group equation (RGE), and these dependencies can be expressed as follows,
\begin{eqnarray}\label{eq:28}
\notag
m_{c}(\mu ) &&= m_{c}(m_{c})\left[\frac{\alpha _s(\mu )}{\alpha _s(m_{c})}\right]^{\frac{12}{33 - 2n_f}}\\
\notag
m_{s}(\mu ) &&= m_{s}(2\mathrm{GeV})\left[\frac{\alpha _s(\mu )}{\alpha _s(2\mathrm{GeV})}\right]^{\frac{12}{33 - 2n_f}}\\
\notag
\left\langle \bar qq \right\rangle (\mu ) &&= \left\langle \bar qq \right\rangle (1\mathrm{GeV})\left[\frac{\alpha _s(1\mathrm{GeV})}{\alpha _s(\mu )}\right]^{\frac{12}{33 - 2n_f}}\\
\notag
\left\langle \bar qg_s\sigma Gq \right\rangle (\mu ) &&= \left\langle \bar qg_s\sigma Gq \right\rangle (1\mathrm{GeV})\left[\frac{\alpha _s(1\mathrm{GeV})}{\alpha _s(\mu )}\right]^{\frac{2}{33 - 2n_f}}\\
\notag
\left\langle \bar sg_s\sigma Gs \right\rangle (\mu ) &&= \left\langle \bar sg_s\sigma Gs \right\rangle (1\mathrm{GeV})\left[\frac{\alpha _s(1\mathrm{GeV})}{\alpha _s(\mu )}\right]^{\frac{2}{33 - 2n_f}}\\
\notag
\alpha _s(\mu ) &&= \frac{1}{b_0t}\left[ 1 - \frac{b_1}{b_0^2}\frac{\log t}{t} \right.\\
&&\left. + \frac{b_1^2(\log ^2t - \log t - 1) + b_0b_2}{b_0^4t^2} \right]
\end{eqnarray}
where $t=\log\frac{\mu^2}{\Lambda_{QCD}^2}$, $b_0=\frac{33-2n_f}{12\pi}$, $b_1=\frac{153-19n_f}{24\pi^2}$, $b_2=\frac{2857-\frac{5033}{9}n_f+\frac{325}{27}n_f^2}{128\pi^3}$, $\Lambda_{\mathrm{QCD}}$ is taken to be $292$ MeV for the quark flavor $n_f=4$ in the present work~\cite{ParticleDataGroup:2024cfk}. The minimum subtraction masses of $c$ and $s$ quarks are taken to be the standard values from PDG~\cite{ParticleDataGroup:2024cfk}, which are $m_c(m_c)=1.275\pm0.025$ GeV and $m_s(\mu=2\mathrm{GeV})=0.095\pm0.005$ GeV. As for the energy scale, a typical value of $\mu=1$ GeV is adopted for charmed baryon~\cite{Wang:2010vn}.

To obtain reliable results, an appropriate working region about Borel parameters should be selected. In this region, the values of form factors should have a weak dependence on the Borel parameters. This working region is commonly named as 'Borel platform'. At the same time, two conditions should also be satisfied which are the pole dominance and the convergence of OPE. The pole contribution is defined as,
\begin{eqnarray}\label{eq:29}
\mathrm{Pole}=\frac{\int_{s_{min}}^{s_{0}} ds\int_{u_{min}}^{u_{0}} du\rho^{\mathrm{QCD}}\left(s,u,q^{2}\right)\mathrm{exp}\left(-\frac{s}{\mathrm{M}_{1}^{2}}-\frac{u}{\mathrm{M}_{2}^{2}}\right)}{\int_{s_{min}}^{\infty} ds\int_{u_{min}}^{\infty} du\rho^{\mathrm{QCD}}\left(s,u,q^{2}\right)\mathrm{exp}\left(-\frac{s}{\mathrm{M}_{1}^{2}}-\frac{u}{\mathrm{M}_{2}^{2}}\right)}
\end{eqnarray}
\begin{figure}[htbp] \centering
{\includegraphics[width=0.5\textwidth]{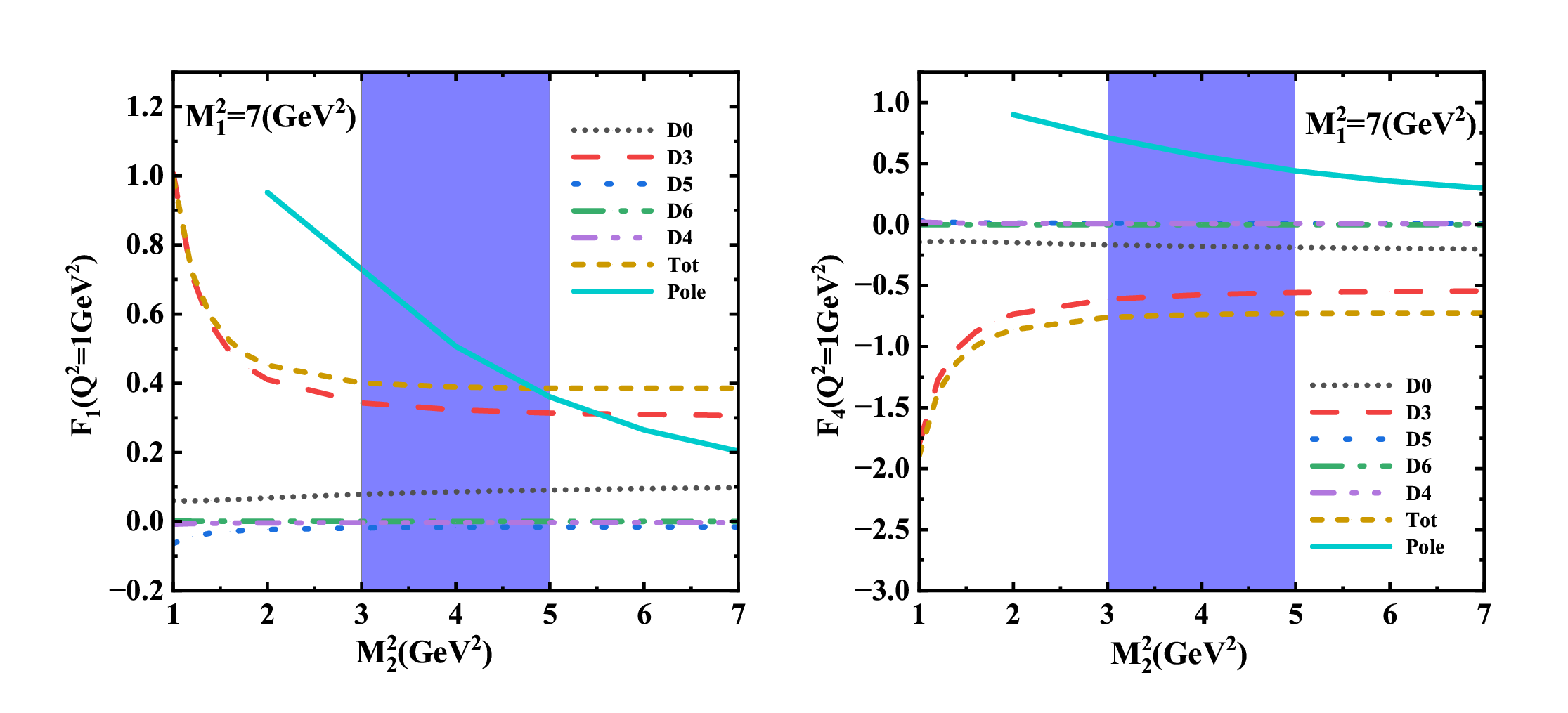}}
\caption{Variations of the pole contributions and contributions of perturbative term, different vacuum condensates with respect to the Borel parameter $\mathrm{M}_{2}^{2}$. These results are for the form factors $F_{1}$ and $F_{4}$ of transition $\Xi_{cc}^{++}\rightarrow\Sigma_{c}^{*+}$. The blue bounds denote the Borel platform.}
\label{pole}
\end{figure}
The pole dominance requires that the pole contribution should be larger than $40\%$, and the convergence of OPE requires the contribution of high dimension condensate terms should be as small as possible. For the form factors $F_{1}$ and $F_{4}$ of transition $\Xi_{cc}^{++}\rightarrow\Sigma_{c}^{*+}$ as examples, we introduce how the Borel platform is determined. Fixing $Q^{2}=1$ GeV$^{2}$, we plot the pole contribution and contributions of perturbative part and different vacuum condensate terms in Fig. \ref{pole} which explicitly illustrates the dependence of the results on the Borel parameters. From this figure, we can see that the pole contribution (Pole) decreases with the increase of Borel parameters. In addition, we can also see that main contributions come from quark condensate (D3) and purturbative part (D0), and more larger values of the Borel parameters are taken more smaller contributions of vacuum condensate terms become. That is to say, the condition of convergence of OPE can be well satisfied when the larger Borel parameters are taken. After repeated trial and contrast, the Borel platform for M$_{2}^{2}$ is taken to be $3\sim5$ GeV$^2$ marked as blue area in Fig. \ref{pole}. In this region, the average value of pole contribution is about $50\%$ and at the same time the contributions of different dimensions satisfy the relation,
$\mathrm{D}3>\mathrm{D}0>\mathrm{D}5\gg\mathrm{D}4>\mathrm{D}6$.
The contributions originate from high dimension condensates $\left\langle {g_s^2GG} \right\rangle$ and $g_{s}^{2}\langle \overline{q}q\rangle^{2}$ are both smaller than $2\%$.
\begin{figure}[htbp] \centering
\begin{subfigure}{\includegraphics[width=0.5\textwidth]{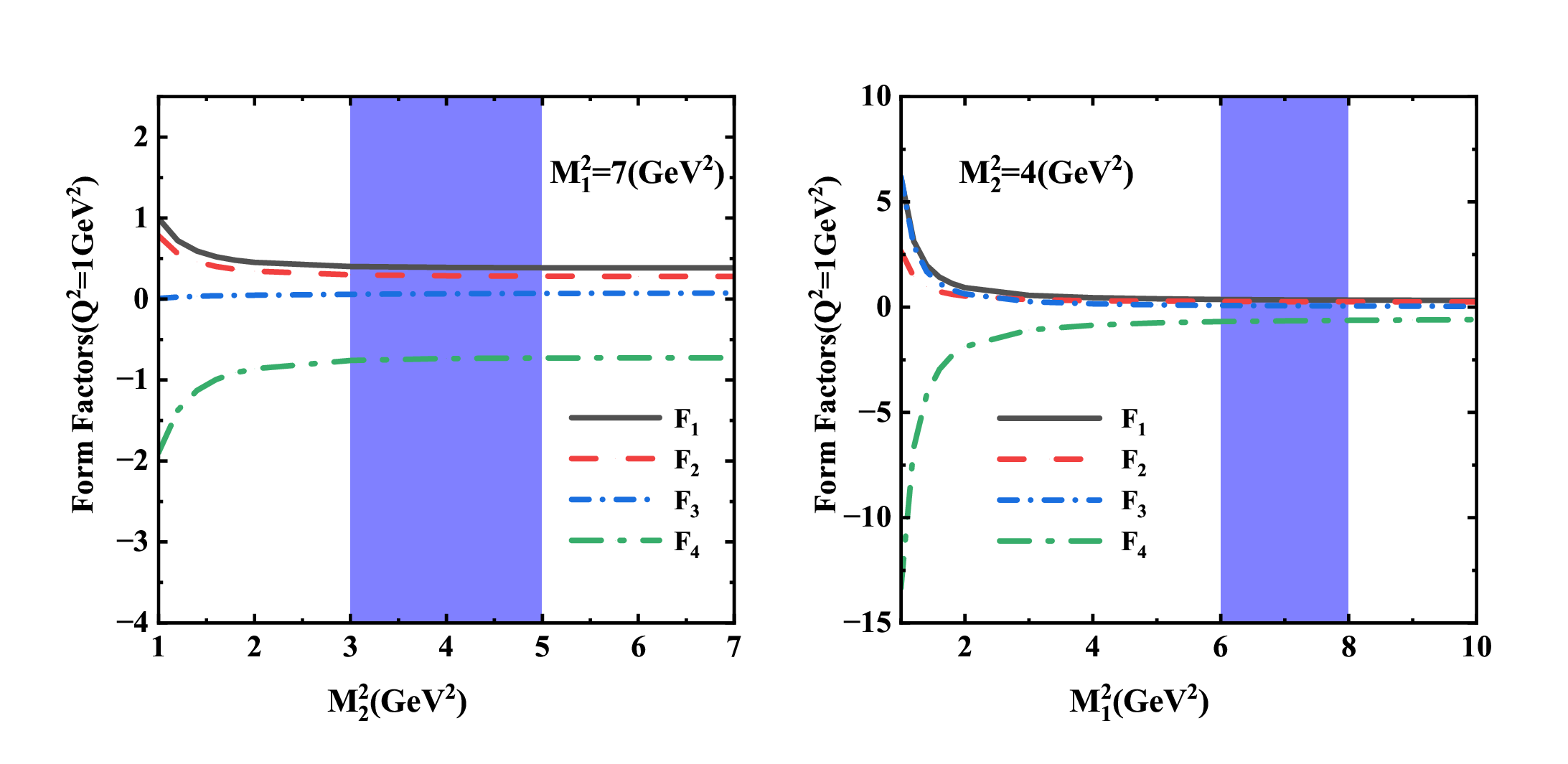}}
\end{subfigure}
\begin{subfigure}{\includegraphics[width=0.5\textwidth]{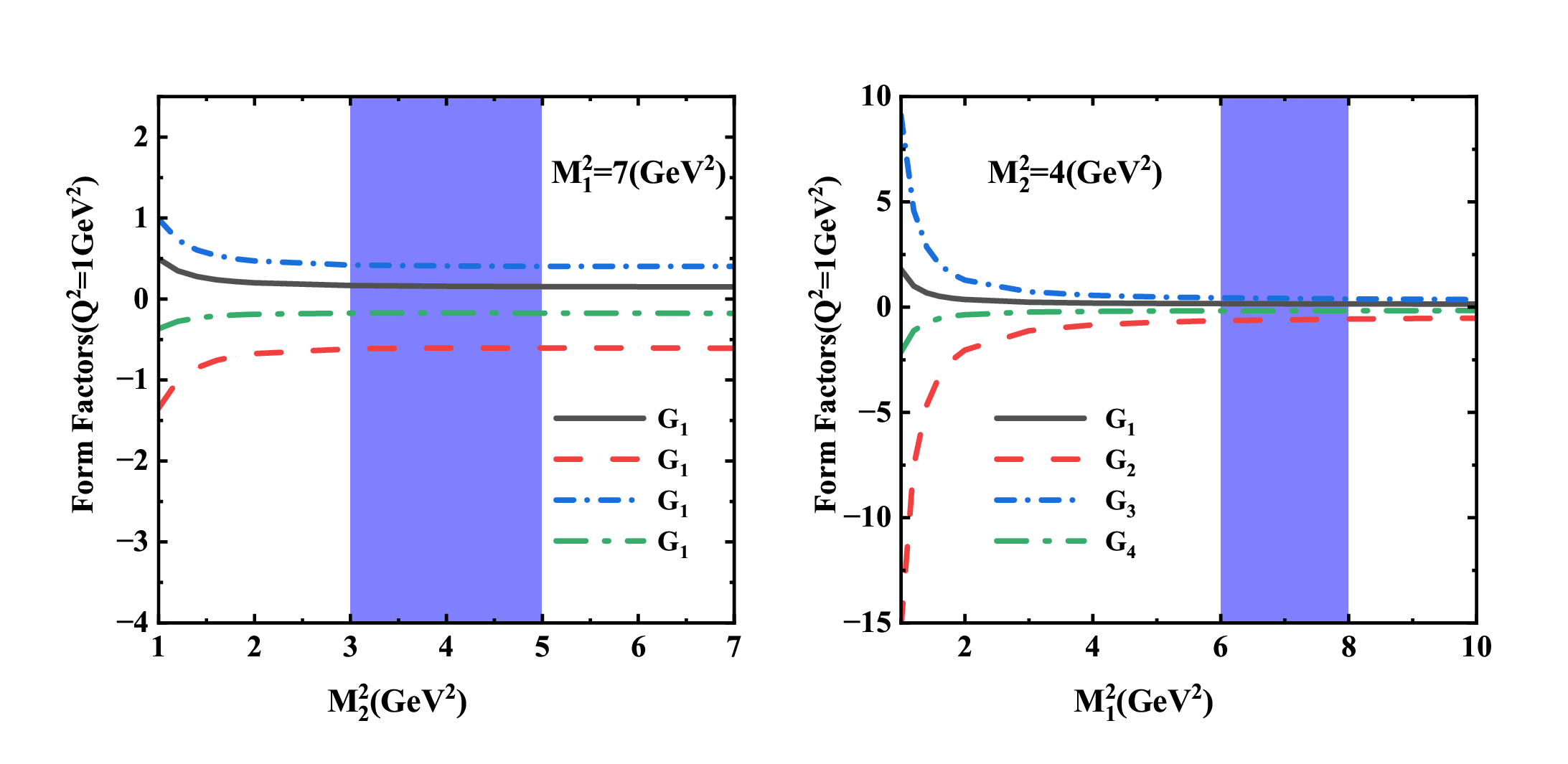}}
\caption{Variations of the form factors ($\Xi_{cc}^{++}\rightarrow\Sigma_{c}^{*+}$ transition process) with respect to the Borel parameters $\mathrm{M}_{1}^{2}$ and $\mathrm{M}_{2}^{2}$, where the blue bounds denote the Borel platform.}
\label{borel}
\end{subfigure}
\end{figure}

Another two parameters to be selected are the threshold parameters $s_0$ and $u_0$ which are used to eliminate the contributions of excited and continuum states. Their values are usually taken to be $s_0=(m_{\mathcal{B}_{1}}+\Delta_{1})^{2}$ and $u_0=(m_{\mathcal{B}_{2}^{*}}+\Delta_{2})^{2}$. Commonly, the energy gap between the ground state and the first excited one of heavy baryons is about $500$ MeV, thus the value of $\Delta_{1}=\Delta_{2}=0.5$ GeV are taken in the present work. For the transition precess $\Xi_{cc}^{++}\rightarrow\Sigma_{c}^{*+}$ as an example, we plot variations of the form factors with respect to the Borel parameters $\mathrm{M}_{1}^{2}$ and $\mathrm{M}_{2}^{2}$ in Fig. \ref{borel}. From this figure, we can see that the results show well stability and weak dependence on the Borel parameters in the Borel platform.

\begin{figure}[htbp]
	\centering
	\includegraphics[width=8.5cm]{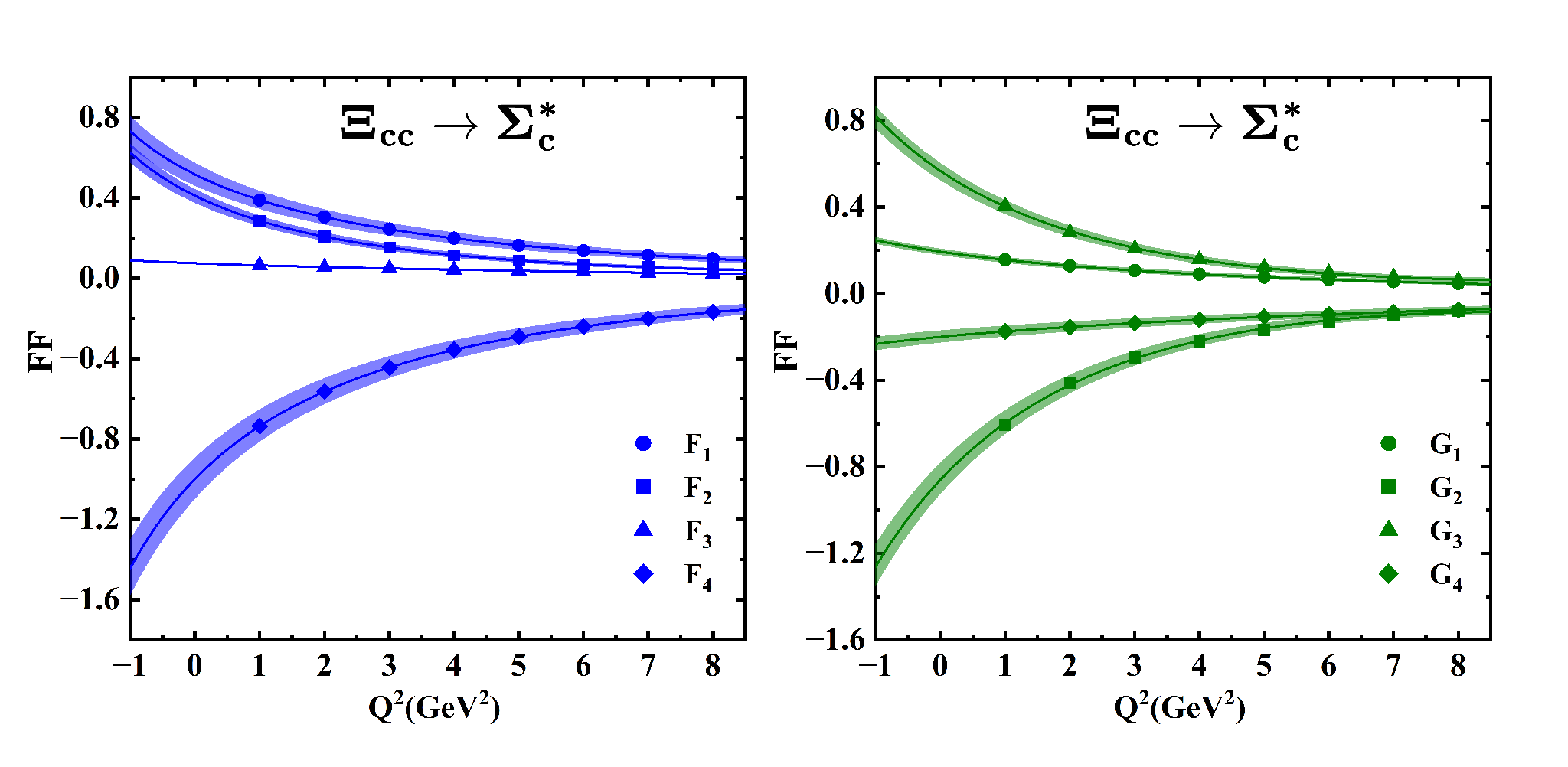}
	\centering
	\includegraphics[width=8.5cm]{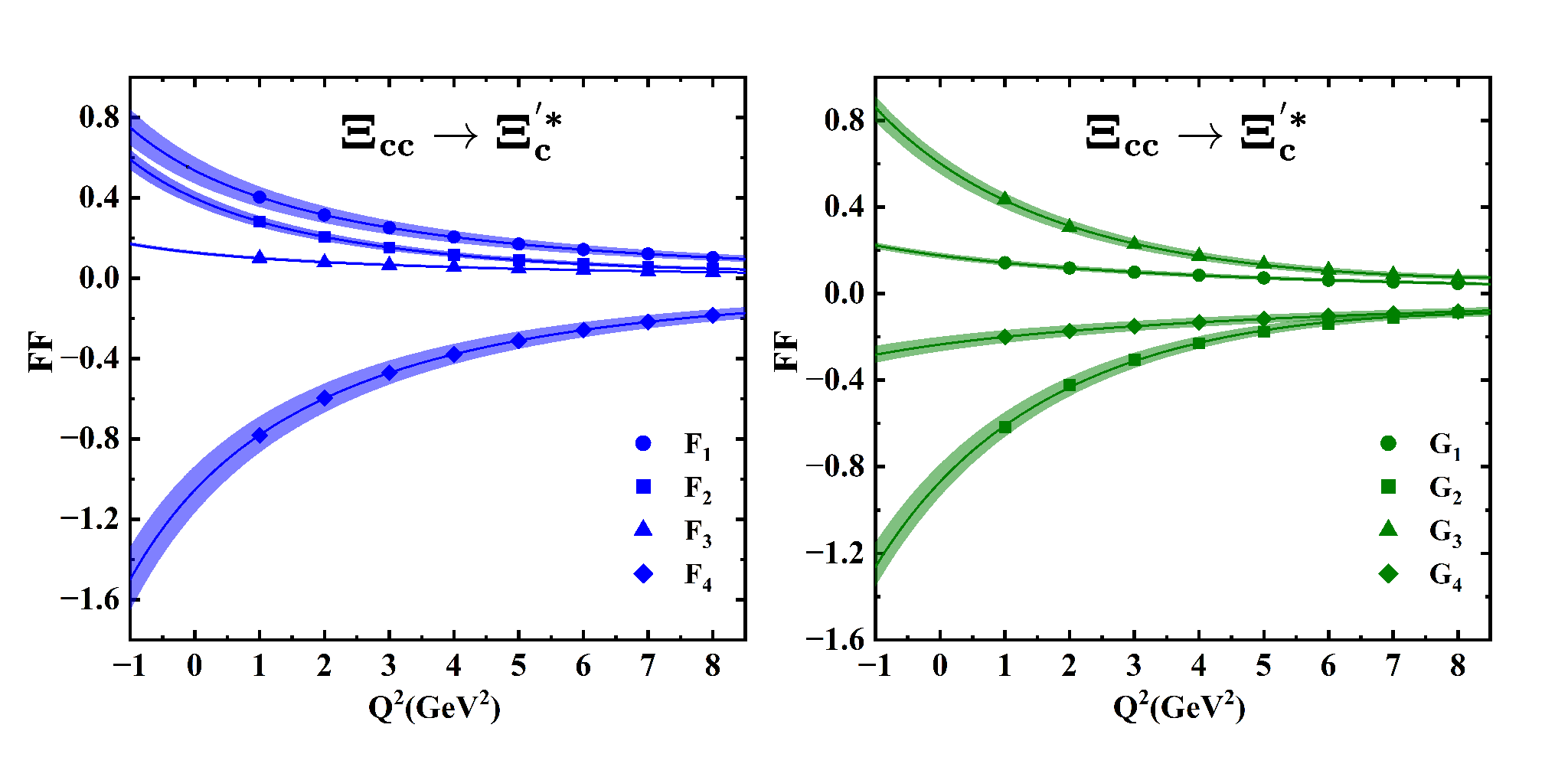}
	\caption{The fitting results of the form factors for transition processes $\Xi_{cc}^{++}\rightarrow\Sigma_{c}^{*+}$ and $\Xi_{cc}^{++}\rightarrow\Xi_{c}^{\prime*+}$.}
\label{fitting1}
\end{figure}
\begin{figure}[htbp]
	\centering
	\includegraphics[width=8.5cm]{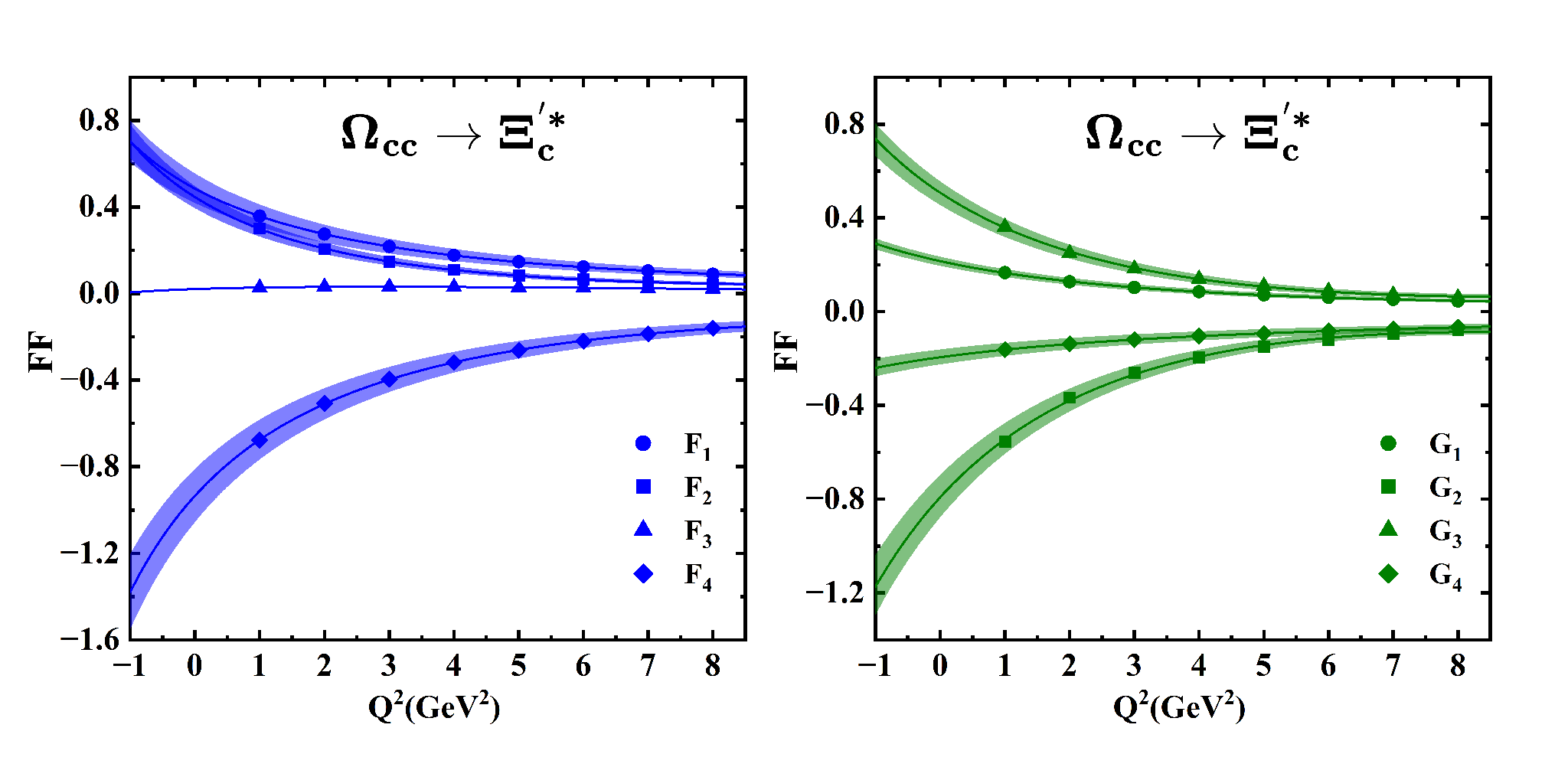}
	\centering
	\includegraphics[width=8.5cm]{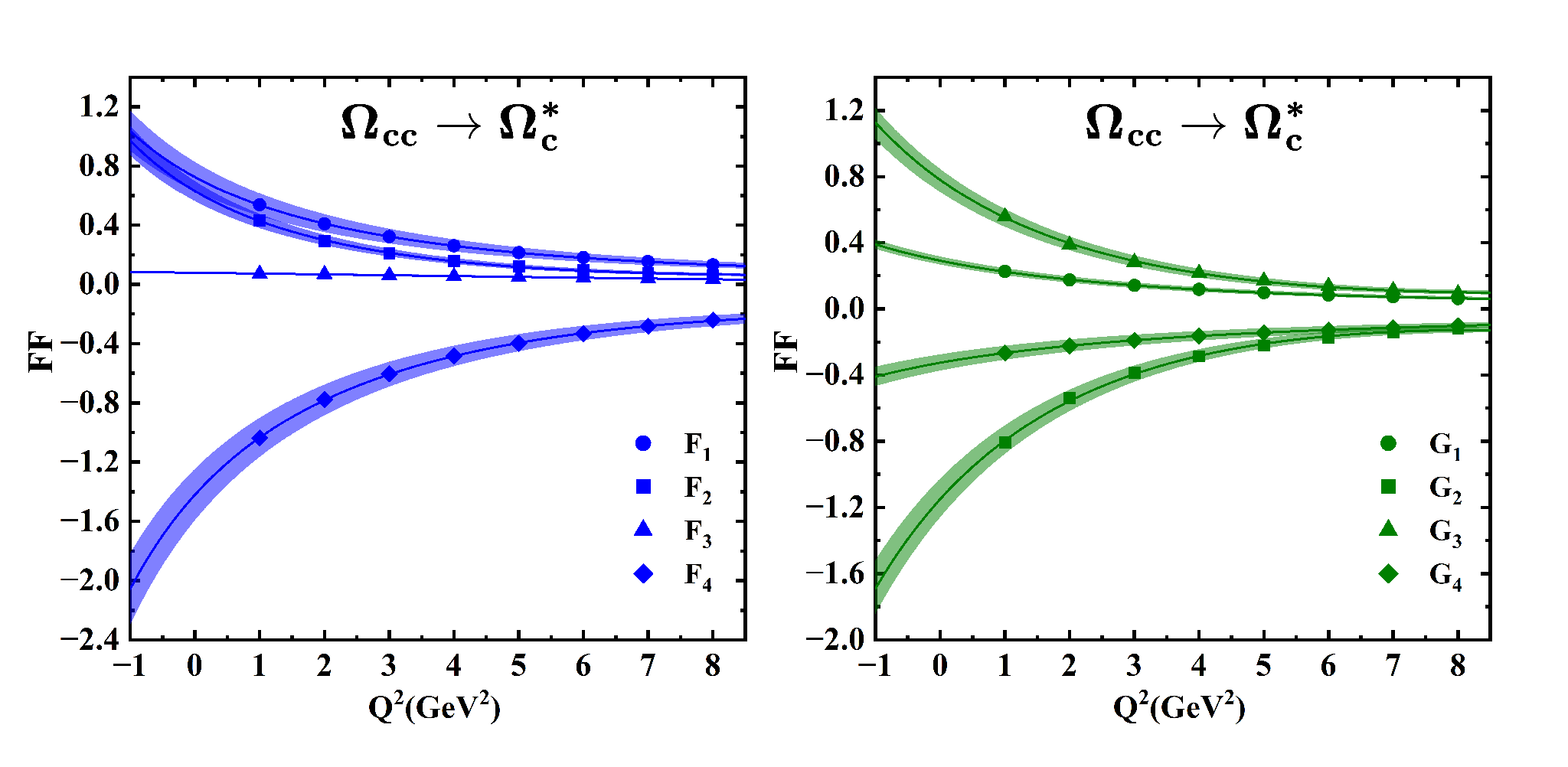}
	\caption{The fitting results of the form factors for transition processes $\Omega_{cc}^{+}\rightarrow\Xi_{c}^{\prime*0}$ and $\Omega_{cc}^{+}\rightarrow\Omega_{c}^{*0}$.}
\label{fitting2}
\end{figure}
After all of the conditions of QCDSR are satisfied, we can obtain the momentum dependent form factors $F_{i}(Q^2)$ and $G_{i}(Q^2)$ by taking different values of $Q^2$ in space-like region ($Q^2>0$). The results are explicitly shown in Figs. \ref{fitting1} and \ref{fitting2}, where $Q^2$ are in the range of $1\sim8$ GeV$^2$. To obtain the values of the form factors in time-like region ($Q^{2}<0$), it is necessary to fit the numerical results with appropriate analytical function and then extrapolate them into time-like region. The z series expand approach is commonly employed to fit the form factors \cite{Boyd:1994tt}. With this method, the form factors can be expressed as,
\begin{eqnarray}\label{eq:30}
\notag
&&F_{i}(Q^2) = \\ \notag
&&\frac{F(0)}{1 + Q^2/m_{\mathrm{pole}}^{2}}\times\Bigg\{1+a\left[z(Q^{2})-z(0)-\frac{1}{3}[z(Q^2)^{3}-z(0)^3]\right]\\ \notag
&&+b\left[z(Q^{2})^{2}-z(0)^{2}-\frac{2}{3}[z(Q^2)^{3}-z(0)^3]\right]\Bigg\} \\
\notag
&&G_{i}(Q^2) = \\ \notag
&&\frac{G(0)}{1 + Q^2/m_{\mathrm{pole}}^{2}}\times\Bigg\{1+c\left[z(Q^{2})-z(0)-\frac{1}{3}[z(Q^2)^{3}-z(0)^3]\right]\\
&&+d\left[z(Q^{2})^{2}-z(0)^{2}-\frac{2}{3}[z(Q^2)^{3}-z(0)^3]\right]\Bigg\}
\end{eqnarray}
with $z(Q^{2})=\frac{\sqrt{t_{+}^{2}+Q^{2}}-\sqrt{t_{+}^{2}-t_{-}^{2}}}{\sqrt{t_{+}^{2}+Q^{2}}+\sqrt{t_{+}^{2}-t_{-}^{2}}}$ and $t_{\pm}=(m_{\mathcal{B}_{1}}\pm m_{\mathcal{\mathcal{B}}_{2}^{*}})^{2}$.
In these equations, $a\sim d$ are fitting parameters and their values are all listed in Tab. \ref{FP}. The fitting curves of the form factors are explicitly shown in Figs. \ref{fitting1} and \ref{fitting2}. From these figures, we can see that the form factors are fitted well by z series expand approach. Thus, it is reliable for us to obtain the values of form factors in time-like region which can be used to analyze the semileptonic decays of $\Xi_{cc}$ and $\Omega_{cc}$ baryons.

\begin{table*}[htbp]
\caption{The values of form factors (FF) at $Q^2=0$ and fitting parameters of different form factors by z series expand approach.}
\begin{ruledtabular}
\renewcommand{\arraystretch}{1.3}
\label{FP}
\begin{tabular}{c c c c c c c| c c c c c c}
				Mode&FF&$F_{i}(0)$&$a$&$b$&$F(q^{2}_{max})$&$F_{i}(0)$\cite{Hu:2020mxk}&FF&$G_{i}(0)$&$c$&$d$&$G(q^{2}_{max})$&$G_{i}(0)$\cite{Hu:2020mxk} \\
				\hline
				\multirow{4}{*}				{$\Xi_{cc}^{++}\rightarrow\Sigma_{c}^{*+}$}&$F_{1}$&$0.52_{-0.06}^{+0.06}$&$-9.45_{-1.19}^{+0.97}$&$2.39_{-4.12}^{+5.16}$&$0.75_{-0.09}^{+0.09}$&-0.979&
				$G_{1}$&$0.19_{-0.02}^{+0.01}$&$-8.42_{-2.28}^{+2.09}$&$-5.77_{-14.26}^{+16.11}$&$0.24_{-0.03}^{+0.01}$&-5.792\\
			&$F_{2}$&$0.41_{-0.04}^{+0.03}$&$-23.25_{-1.52}^{+1.50}$&$134.38_{-14.71}^{+14.11}$&$0.59_{-0.06}^{+0.04}$&0.66&
			$G_{2}$&$-0.86_{-0.07}^{+0.08}$&$-35.05_{-1.49}^{+1.33}$&$286.90_{-14.34}^{+16.15}$&$-1.09_{-0.09}^{+0.10}$&3.84\\
			&$F_{3}$&$0.07_{-0.004}^{+0.004}$&$17.64_{-4.35}^{+4.35}$&$-285.62_{-36.18}^{+33.18}$&$0.10_{-0.006}^{+0.006}$&0.42&
			$G_{3}$&$0.57_{-0.04}^{+0.04}$&$-31.94_{-1.73}^{+1.57}$&$251.65_{-16.57}^{+18.25}$&$0.72_{-0.05}^{+0.05}$&2.28\\
			&$F_{4}$&$-1.00_{-0.10}^{+0.10}$&$-13.50_{-1.08}^{+0.98}$&$47.38_{-5.32}^{+4.95}$&$-1.44_{-0.14}^{+0.14}$&-1.969&
			$G_{4}$&$-0.20_{-0.03}^{+0.03}$&$6.25_{-0.47}^{+0.86}$&$-129.65_{-8.55}^{+1.37}$&$-0.25_{-0.04}^{+0.04}$&0.393\\
			\hline
			\multirow{4}{*}			 {$\Xi_{cc}^{++}\rightarrow\Xi_{c}^{'*+}$}&$F_{1}$&$0.54_{-0.07}^{+0.07}$&$-13.53_{-0.77}^{+0.57}$&$57.73_{-1.00}^{+2.26}$&$0.69_{-0.09}^{+0.09}$&-1.148&
			 $G_{1}$&$0.18_{-0.02}^{+0.01}$&$-9.13_{-2.17}^{+2.32}$&$13.49_{-19.60}^{+16.19}$&$0.21_{-0.02}^{+0.01}$&-10.350\\
			 &$F_{2}$&$0.40_{-0.04}^{+0.04}$&$-24.51_{-1.30}^{+1.08}$&$164.35_{-10.26}^{+13.01}$&$0.51_{-0.05}^{+0.05}$&0.72&
			 $G_{2}$&$-0.87_{-0.07}^{+0.08}$&$-34.78_{-1.28}^{+1.26}$&$307.93_{-14.61}^{+14.57}$&$-1.04_{-0.08}^{+0.10}$&6.43\\
			 &$F_{3}$&$0.13_{-0.009}^{+0.009}$&$-6.92_{-2.90}^{+2.41}$&$-11.15_{-21.35}^{+27.43}$&$0.17_{-0.01}^{+0.01}$&0.52&
			 $G_{3}$&$0.60_{-0.05}^{+0.04}$&$-31.53_{-1.44}^{+1.42}$&$267.88_{-15.78}^{+15.77}$&$0.72_{-0.06}^{+0.05}$&4.70\\
			 &$F_{4}$&$-1.05_{-0.12}^{+0.12}$&$-16.58_{-0.84}^{+0.79}$&$93.65_{-4.49}^{+4.10}$&$-1.34_{-0.15}^{+0.15}$&-2.297&
			 $G_{4}$&$-0.24_{-0.03}^{+0.03}$&$-0.20_{-0.68}^{+0.82}$&$-62.14_{-0.76}^{-2.57}$&$-0.28_{-0.04}^{+0.04}$&0.404\\
			 \hline			 \multirow{4}{*}{$\Omega_{cc}^{+}\rightarrow\Xi_{c}^{'*0}$}&$F_{1}$&$0.49_{-0.07}^{+0.07}$&$-15.26_{-0.91}^{+0.96}$&$85.80_{-9.14}^{+7.64}$&$0.70_{-0.10}^{+0.10}$&-1.017&
			 $G_{1}$&$0.22_{-0.02}^{+0.02}$&$-20.59_{-2.18}^{+2.09}$&$152.50_{-19.94}^{+20.95}$&$0.28_{-0.03}^{+0.03}$&-6.244\\
			 &$F_{2}$&$0.45_{-0.05}^{+0.05}$&$-33.23_{-1.00}^{+0.96}$&$285.01_{-10.51}^{+10.65}$&$0.65_{-0.07}^{+0.07}$&0.66&
			 $G_{2}$&$-0.79_{-0.09}^{+0.09}$&$-40.54_{-1.19}^{+1.22}$&$386.80_{-16.16}^{+15.42}$&$-1.01_{-0.11}^{+0.11}$&3.86\\
			 &$F_{3}$&$0.02_{-0.001}^{+0.001}$&$144.87_{-17.63}^{+21.10}$&$-1612.00_{-0.25}^{+0.20}$&$0.03_{-0.001}^{+0.001}$&0.47&
			 $G_{3}$&$0.51_{-0.05}^{+0.05}$&$-36.18_{-1.40}^{+1.39}$&$331.61_{-17.45}^{+17.39}$&$0.65_{-0.06}^{+0.06}$&2.73\\
			 &$F_{4}$&$-0.93_{-0.12}^{+0.12}$&$-19.10_{-0.92}^{+1.01}$&$133.04_{-10.71}^{+8.86}$&$-1.34_{-0.17}^{+0.17}$&-2.045&
			 $G_{4}$&$-0.19_{-0.03}^{+0.03}$&$-4.04_{-1.18}^{+1.51}$&$0.03_{-15.49}^{+11.02}$&$-0.24_{-0.04}^{+0.04}$&0.389\\
			 \hline			 \multirow{4}{*}{$\Omega_{cc}^{+}\rightarrow\Omega_{c}^{'*0}$}&$F_{1}$&$0.73_{-0.10}^{+0.10}$&$-18.73_{-1.06}^{+1.25}$&$132.11_{-17.60}^{+12.21}$&$0.94_{-0.13}^{+0.13}$&-1.178&
			 $G_{1}$&$0.29_{-0.02}^{+0.03}$&$-20.49_{-2.66}^{+1.77}$&$160.42_{-17.50}^{+30.94}$&$0.35_{-0.02}^{+0.04}$&-10.670\\
			 &$F_{2}$&$0.63_{-0.07}^{+0.07}$&$-34.84_{-0.96}^{+0.86}$&$328.09_{-10.20}^{+11.61}$&$0.81_{-0.09}^{+0.09}$&0.73&
			 $G_{2}$&$-1.15_{-0.12}^{+0.12}$&$-40.55_{-1.30}^{+1.20}$&$413.34_{-17.28}^{+18.99}$&$-1.38_{-0.14}^{+0.14}$&6.55\\
			 &$F_{3}$&$0.08_{-0.004}^{+0.002}$&$32.05_{-8.90}^{+11.25}$&$-469.48_{-152.33}^{+121.27}$&$0.10_{-0.005}^{+0.002}$&0.54&
			 $G_{3}$&$0.78_{-0.07}^{+0.07}$&$-36.54_{-1.43}^{+1.41}$&$358.31_{-19.52}^{+19.82}$&$0.93_{-0.08}^{+0.08}$&4.85\\
			 &$F_{4}$&$-1.42_{-0.18}^{+0.17}$&$-21.62_{-1.19}^{+0.94}$&$170.48_{-10.99}^{+14.91}$&$-1.82_{-0.23}^{+0.22}$&-2.359&
			 $G_{4}$&$-0.33_{-0.05}^{+0.05}$&$-8.89_{-1.64}^{+1.50}$&$46.77_{-17.36}^{+20.28}$&$-0.39_{-0.06}^{+0.06}$&0.415
\end{tabular}
\end{ruledtabular}
\end{table*}

A similar analysis about the form factors was also carried out in the frame work of light-front quark model in Ref. \cite{Hu:2020mxk}. For comparison, the results of the present work and those obtained with quark model \cite{Hu:2020mxk} are all listed int Tab. \ref{FP}. It is noted that if a factor of $\sqrt{2}$ is multiplied to $F_{i}/G_{i}$ for $\Xi_{cc}^{++}\rightarrow\Sigma_{c}^{*+}$, we can obtain the results for transition $\Xi_{cc}^{++}\rightarrow\Sigma_{c}^{*0}$. Thus, the results of the latter are not listed in the table. It is shown that our results are apparently lower than the values of quark model. Compared with the present work, another difference in Ref. \cite{Hu:2020mxk} is that they employ the analytical function with two pole structure to fit the numerical results of form factors. Their fitting curves illustrate that the absolute values of form factors become lower after being extrapolated into the time-like region. In the present work, the situation is exactly the opposite, where the absolute values increase after being fitted into time-like region (See Figs. \ref{fitting1} and \ref{fitting2}). From Tab. \ref{FP}, we can see that the on-shell values of $F(q_{max}^{2})$/$G(q_{max}^{2})$ , $q_{max}^{2}=(m_{\mathcal{B}_{1}}-m_{\mathcal{B}_{2}^{*}})^{2}$, are larger than the values at zero. In another literature \cite{Lu:2023rmq}, the form factors of $\Xi_{cc}^{++}\rightarrow\Sigma_{c}^{*+}$ were also analyzed by the light-front quark model. The results of $F_{1}(0)\sim F_{4}(0)$ and $G_{1}(0)\sim G_{4}(0)$ are ($-0.0179,0.374,-0.381,0.429$) and ($-0.453,0,0.320,-1.14$), which are inconsistent with those of the present work and Ref. \cite{Hu:2020mxk}.

\section{Semileptonic decays of $\Xi_{cc}$ and $\Omega_{cc}$ baryons}\label{sec4}

As a phenomenological application, the results of form factors are applied to analyze the decay widths and branching ratios of semileptonic decays $\Xi_{cc}^{++}\rightarrow \Sigma_{c}^{*+}l^{+}\nu_{l}$, $\Xi_{cc}^{++}\rightarrow \Xi_{c}^{\prime*+}l^{+}\nu_{l}$, $\Omega_{cc}^{+}\rightarrow\Xi_{c}^{\prime*0}l^{+}\nu_{l}$ and $\Omega_{cc}^{+}\rightarrow \Omega_{c}^{*0}l^{+}\nu_{l}$.
The effective Hamiltonian of these processes can be expressed as,
\begin{eqnarray}\label{eq:31}
\notag
H_{eff} =&&\frac{G_F}{\sqrt 2 }\Big(V_{cs}^{*}\bar s\gamma _\mu (1 - \gamma _5)c\bar{ v}_l\gamma _\mu (1 - \gamma _5)l \\
&&+V_{cd}^{*}\bar d\gamma _\mu (1 - \gamma _5)c\bar{ v}_l\gamma _\mu (1 - \gamma _5)l\Big)
\end{eqnarray}
where $G_F$ is the Fermi constant and $V_{cs}^{*}$ and $V_{cd}^{*}$ are the Cabibbo$-$Kobayashi$-$Maskawa (CKM) matrix elements. Their values are taken from the Particle Data Group \cite{ParticleDataGroup:2024cfk},
\begin{eqnarray}
\notag
G_{F}=1.166\times10^{-5}\mathrm{GeV}^{-2},|V_{cd}|=0.225,|V_{cs}|=0.974
\end{eqnarray} With this Hamiltonian, the transition matrix element of the semileptoic decay processes $\Xi_{cc}^{++}\rightarrow\Sigma_{c}^{*+}$, $\Xi_{cc}^{++}\rightarrow\Xi_{c}^{\prime*+}$, $\Omega_{cc}^{+}\rightarrow\Xi_{c}^{\prime*0}$ and $\Omega_{cc}^{+}\rightarrow\Omega_{c}^{*0}$ can be expressed as,
\begin{eqnarray}\label{eq:32}
\notag
T &&= \left\langle \mathcal{B}_{2}^{*}\right|H_{eff}\left| \mathcal{B}_{1} \right\rangle \\
\notag
&&=\frac{G_F}{\sqrt 2}V_{cq^{\prime}}^{*}\left\langle \mathcal{B}_{2}^{*}\right|\bar q^{\prime}\gamma _\mu (1 - \gamma _5)c\left| \mathcal{B}_{1} \right\rangle\\
&&\times \left\langle l\bar{\nu}_l\right|\bar v_l\gamma _\mu(1 - \gamma _5)l\left| 0 \right\rangle,
\end{eqnarray}
The leptonic matrix element can be formulated as the following form by electroweak perturbation theory,
\begin{eqnarray}\label{eq:33}
\langle l\bar{\nu}_l |\bar v_l\gamma _\mu(1 - \gamma _5)l | 0 \rangle  = \bar u_{v,s}\gamma _\mu(1 - \gamma _5)u_{l,s'},
\end{eqnarray}
Here, $u_{v,s}$ and $u_{l,s'}$ are the spinor wave functions of $\bar{\nu}_l$ and $l$, and the subscripts $v(l)$ and $s(s')$ represent the four momentum and spin.
The helicity amplitudes will be used in the calculation and for the vector current and the axial-vector current, they are written as,
\begin{eqnarray}\label{eq:34}
\notag
H_{\frac{1}{2}t}^{V}=&&-\sqrt{\frac{2}{3}}\alpha^{V}_{\frac{1}{2}t}(\omega-1)\Big[f_{1}^{V}(q^{2})m_{\mathcal{B}_{1}}-f_{2}^{V}(q^{2})m_{+}
\\ \notag
&&+f_{3}^{V}(q^{2})\frac{m_{\mathcal{B}_{2}^{*}}}{m_{\mathcal{B}_{1}}}(m_{\mathcal{B}_{1}}\omega-m_{\mathcal{B}_{2}^{*}})+f_{4}^{V}(q^{2})\frac{q^{2}}{m_{\mathcal{B}_{1}}}\Big] \\
\notag
H_{\frac{1}{2}0}^{V}=&&-\sqrt{\frac{2}{3}}\alpha^{V}_{\frac{1}{2}0}\Big[f_{1}^{V}(q^{2})(m_{\mathcal{B}_{1}}\omega-m_{\mathcal{B}_{2}^{*}})
-f_{2}^{V}(q^{2})(\omega+1)m_{-}\\ \notag
&&+f_{3}^{V}(q^{2})(\omega^{2}-1)m_{\mathcal{B}_{2}^{*}}\Big]\\ \notag
H_{\frac{1}{2}1}^{V}=&&\sqrt{\frac{1}{6}}\alpha^{V}_{\frac{1}{2}1}\Big[f_{1}^{V}(q^{2})-2f_{2}^{V}(q^{2})(\omega+1)\Big]\\ \notag
H_{\frac{3}{2}1}^{V}=&&-\frac{1}{\sqrt{2}}\alpha^{V}_{\frac{1}{2}1}f_{1}^{V}(q^{2})\\
\notag
H_{\frac{1}{2}t}^{A}=&&\sqrt{\frac{2}{3}}\alpha^{A}_{\frac{1}{2}t}(\omega+1)\Big[f_{1}^{A}(q^{2})m_{\mathcal{B}_{1}}-f_{2}^{A}(q^{2})m_{-}
\\ \notag
&&+f_{3}^{A}(q^{2})\frac{m_{\mathcal{B}_{2}^{*}}}{m_{\mathcal{B}_{1}}}(m_{\mathcal{B}_{1}}\omega-m_{\mathcal{B}_{2}^{*}})+f_{4}^{A}(q^{2})\frac{q^{2}}{m_{\mathcal{B}_{1}}}\Big] \\
\notag
H_{\frac{1}{2}0}^{A}=&&\sqrt{\frac{2}{3}}\alpha^{A}_{\frac{1}{2}0}\Big[f_{1}^{A}(q^{2})(m_{\mathcal{B}_{1}}\omega-m_{\mathcal{B}_{2}^{*}})
-f_{2}^{A}(q^{2})(\omega-1)m_{+}\\ \notag
&&+f_{3}^{A}(q^{2})(\omega^{2}-1)m_{\mathcal{B}_{2}^{*}}\Big]\\ \notag
H_{\frac{1}{2}1}^{A}=&&\sqrt{\frac{1}{6}}\alpha^{A}_{\frac{1}{2}1}\Big[f_{1}^{A}(q^{2})-2f_{2}^{A}(q^{2})(\omega-1)\Big]\\
H_{\frac{3}{2}1}^{A}=&&\frac{1}{\sqrt{2}}\alpha^{A}_{\frac{1}{2}1}f_{1}^{A}(q^{2})
\end{eqnarray}
where
\begin{eqnarray}
\notag
&&\alpha_{\frac{1}{2}t}^{V}=\alpha_{\frac{1}{2}0}^{A}=\sqrt{\frac{2m_{\mathcal{B}_{1}}m_{\mathcal{B}_{2}^{*}}(\omega+1)}{q^{2}}}\\ \notag
&&\alpha_{\frac{1}{2}0}^{V}=\alpha_{\frac{1}{2}t}^{A}=\sqrt{\frac{2m_{\mathcal{B}_{1}}m_{\mathcal{B}_{2}^{*}}(\omega-1)}{q^{2}}}\\ \notag
&&
\alpha_{\frac{1}{2}1}^{V}=2\sqrt{m_{\mathcal{B}_{1}}m_{\mathcal{B}_{2}^{*}}(\omega-1)}\\ \notag
&&
\alpha_{\frac{1}{2}1}^{A}=2\sqrt{m_{\mathcal{B}_{1}}m_{\mathcal{B}_{2}^{*}}(\omega+1)}
\end{eqnarray}
and $\omega=\frac{m_{\mathcal{B}_{1}}^{2}+m_{\mathcal{B}_{2}^{*}}^{2}-q^{2}}{2m_{\mathcal{B}_{1}}m_{\mathcal{B}_{2}^{*}}}$, $m_{\pm}=m_{\mathcal{B}_{1}}\pm m_{\mathcal{B}_{2}^{*}}$ with $m_{\mathcal{B}_{1}}$ and $m_{\mathcal{B}_{2}^{*}}$ being the masses of initial and final baryons. The parametrization about the transition matrix element in the present work (See Eq. (\ref{eq:1})) is different from Eq. (\ref{eq:34}), the relations between these form factors are as follows,
\begin{eqnarray}
\notag
&&f_{1}^{V}(q^{2})=F_{4}^{V}(q^{2}),\quad g_{1}^{V}(q^{2})=G_{4}^{V}(q^{2})\\ \notag
&&f_{2}^{V}(q^{2})=F_{1}^{V}(q^{2}),\quad g_{2}^{V}(q^{2})=G_{1}^{V}(q^{2}) \\ \notag
&&f^{V}_{3}(q^2)=\frac{m_{B_{1}}}{m_{B_{2}^{*}}}F^{V}_{3}(q^2)+F_{2}^{V}(q^{2}) \\ \notag
&&g^{V}_{3}(q^2)=\frac{m_{B_{1}}}{m_{B_{2}^{*}}}G^{V}_{3}(q^2)+G_{2}^{V}(q^{2})\\ \notag
&&f_{4}^{V}(q^{2})=F_{2}^{V}(q^{2}),\quad g_{4}^{V}(q^{2})=G_{2}^{V}(q^{2})
\end{eqnarray}
The amplitudes for negative helicity are given by,
\begin{eqnarray}
\notag
H^{V}_{-\lambda_{2},-\lambda_{W}}=-H^{V}_{\lambda_{2},\lambda_{W}} \quad
H^{A}_{-\lambda_{2},-\lambda_{W}}=H^{A}_{\lambda_{2},\lambda_{W}}
\end{eqnarray}
where $\lambda_{2}$ and $\lambda_{W}$ represent the polarizations of the final baryon and the intermediate $W$ boson, respectively. The helicity amplitude for the $J^{V-A}_{\nu}$ current can can be expressed as,
\begin{eqnarray}\label{eq:35}
H_{\lambda_{2},\lambda_{W}}=H^{V}_{\lambda_{2},\lambda_{W}}-H^{A}_{\lambda_{2},\lambda_{W}}
\end{eqnarray}
Finally, the total amplitude about the decay process $\frac{1}{2}^{+}\rightarrow\frac{3}{2}^{+}$ can be expressed as the combinations of the helicity amplitudes,
\begin{eqnarray}\label{eq:35}
\notag
&&\mathcal{H}_{\frac{1}{2}\rightarrow\frac{3}{2}}=|H_{\frac{1}{2},1}|^{2}+|H_{-\frac{1}{2},-1}|^{2}+|H_{\frac{3}{2},1}|^{2}+|H_{-\frac{3}{2},-1}|^{2}+|H_{\frac{1}{2},0}|^{2}
\\ \notag
&&+|H_{-\frac{1}{2},0}|^{2} +\frac{m_{l}^{2}}{2q^{2}}\Big(3|H_{\frac{1}{2},t}|^{2}+3|H_{-\frac{1}{2},t}|^{2}+|H_{\frac{1}{2},1}|^{2}+|H_{-\frac{1}{2},-1}|^{2}\\
&&+|H_{\frac{3}{2},1}|^{2}+|H_{-\frac{3}{2},-1}|^{2}+|H_{\frac{1}{2},0}|^{2}+|H_{-\frac{1}{2},0}|^{2}\Big)
\end{eqnarray}
According to the helicity amplitudes, differential decay width can be decomposed into the following longitudinally and transversely polarized components,
\begin{figure}[htbp] \centering
\begin{subfigure}{\includegraphics[width=0.48\textwidth]{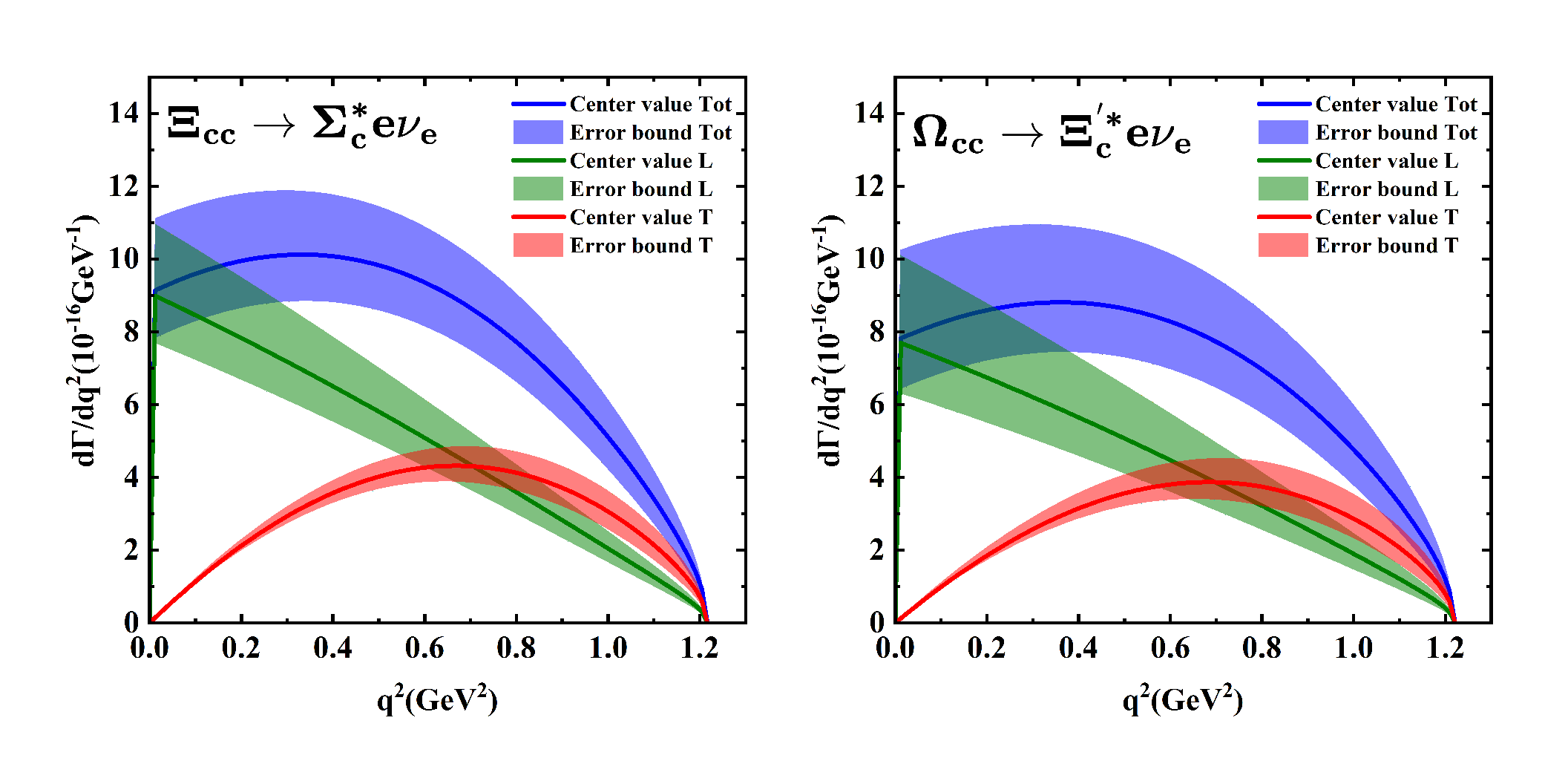}}
\end{subfigure}
\begin{subfigure}{\includegraphics[width=0.48\textwidth]{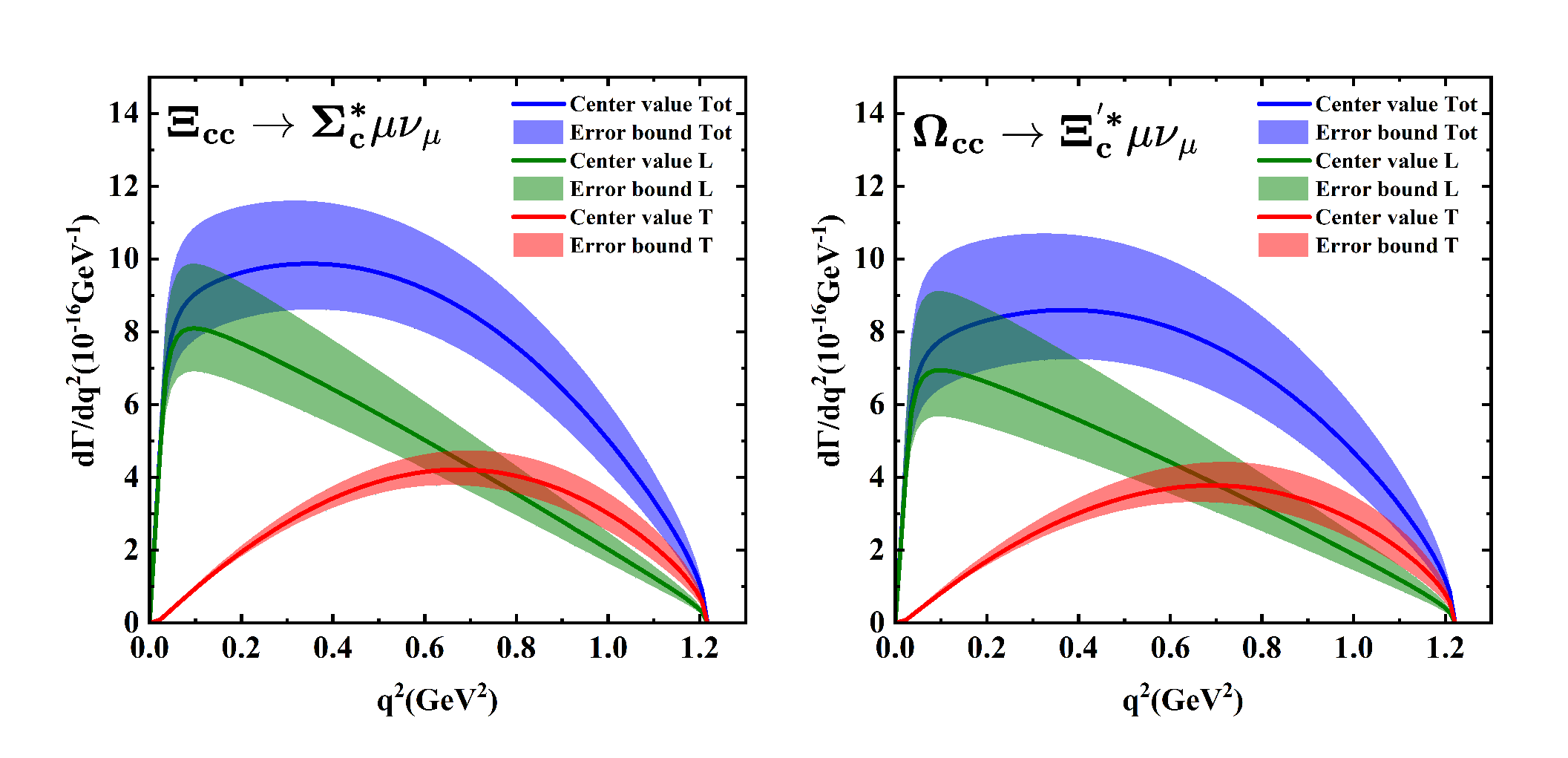}}
\caption{Variations of the differential decay width with respect to $q^2$ for transition processes $\Xi_{cc}^{++}\rightarrow \Sigma_{c}^{*+}l^{+}\nu_{l}$ and $\Xi_{cc}^{++}\rightarrow \Xi_{c}^{\prime*+}l^{+}\nu_{l}$.}
\label{decay1}
\end{subfigure}
\begin{subfigure}{\includegraphics[width=0.48\textwidth]{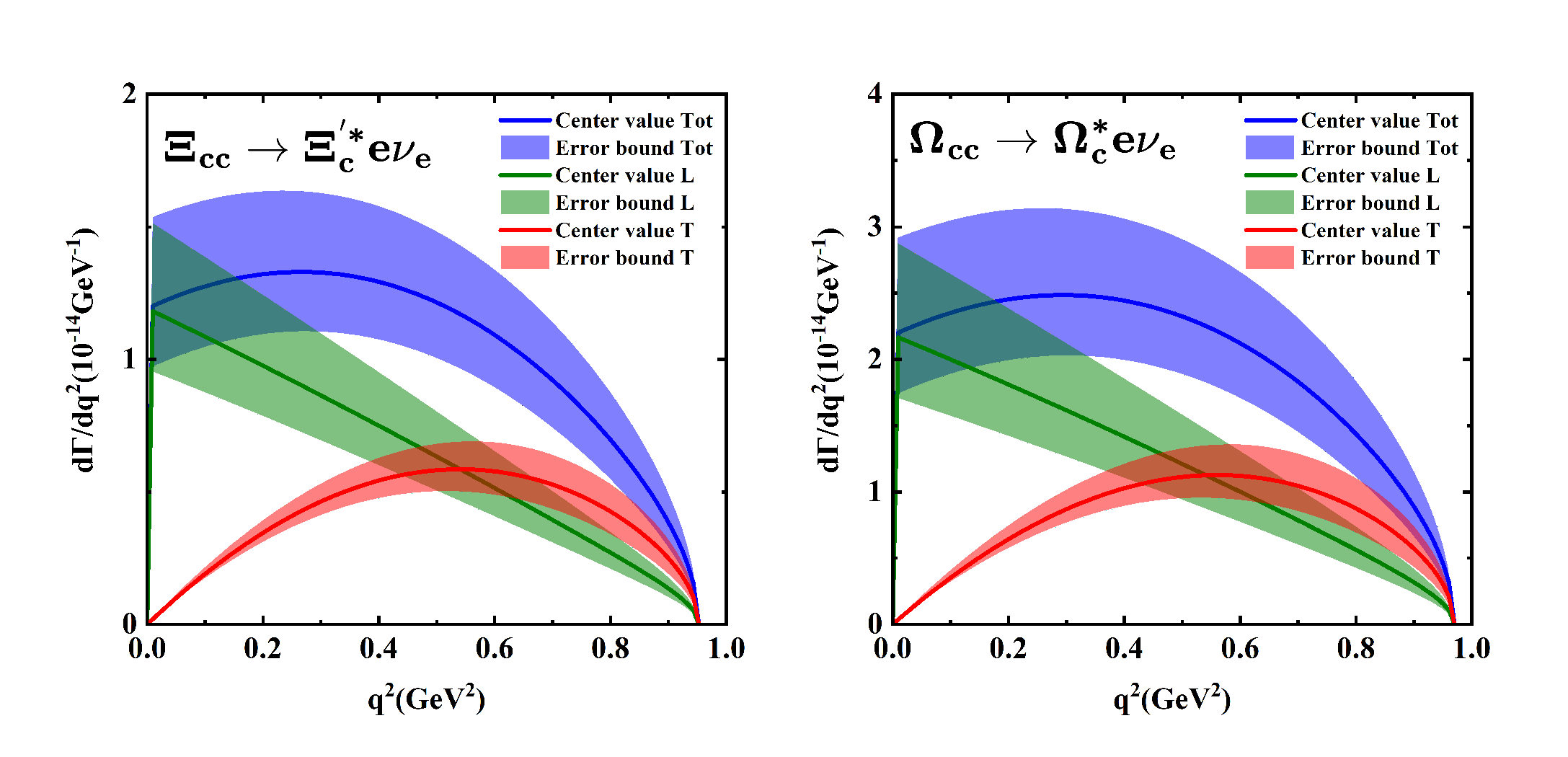}}
\end{subfigure}
\begin{subfigure}{\includegraphics[width=0.48\textwidth]{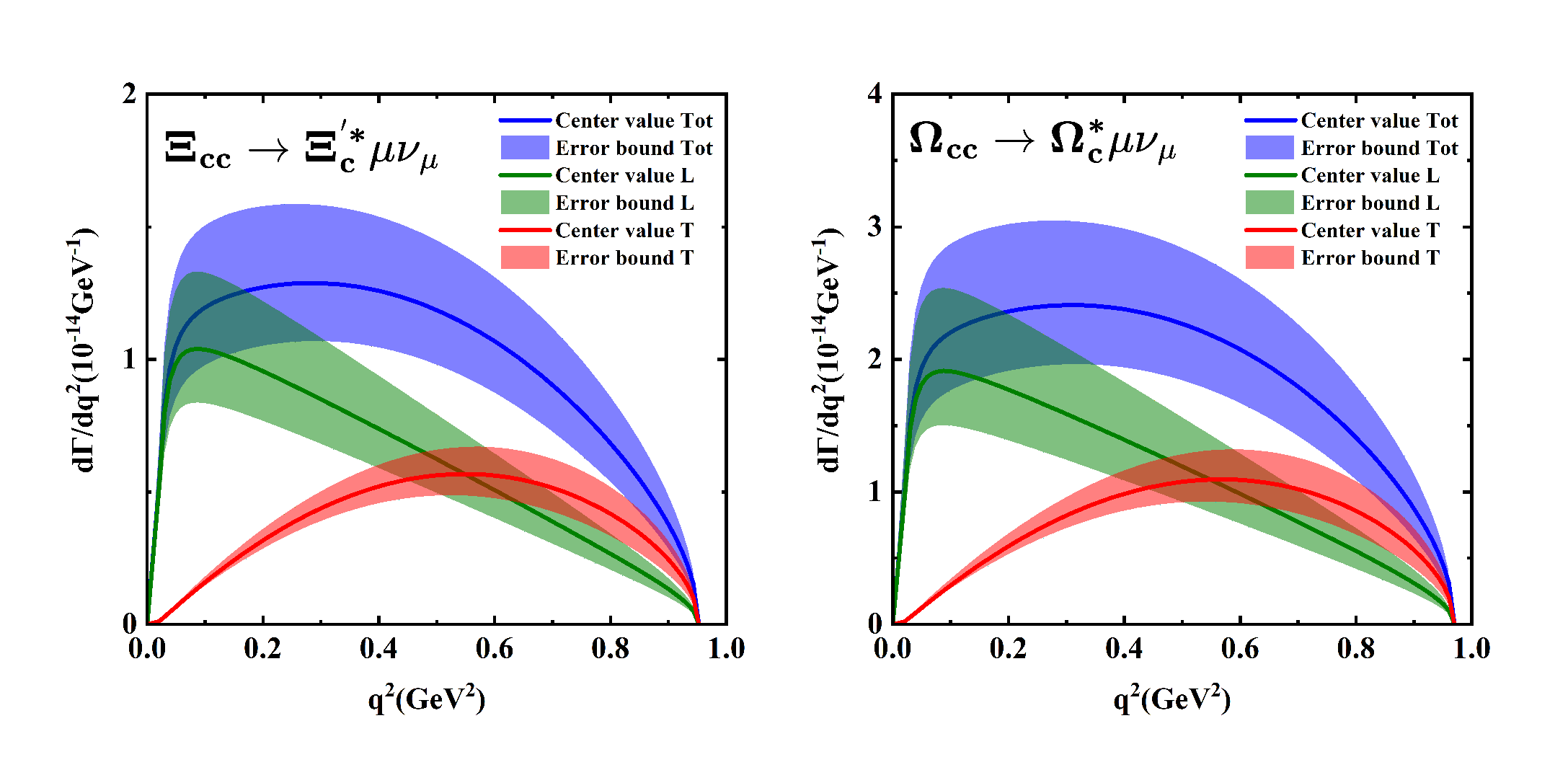}}
\caption{Variations of the differential decay width with respect to $q^2$ for transition processes $\Omega_{cc}^{+}\rightarrow \Xi_{c}^{\prime*0}l^{+}\nu_{l}$ and $\Omega_{cc}^{+}\rightarrow \Omega_{c}^{*0}l^{+}\nu_{l}$.}
\label{decay2}
\end{subfigure}
\end{figure}
\begin{eqnarray}\label{eq:36}
\notag
&&\frac{d\Gamma_{L}}{dq^{2}}=\frac{G_{F}^{2}|V_{CKM}|^{2}}{192\pi^{3}}\frac{m_{\mathcal{B}_{2}^{*}}}{m_{\mathcal{B}_{1}}^{2}}\frac{(q^{2}-m_{l}^{2})^{2}}{q^{2}}
\sqrt{\omega^2-1}\Bigg[|H_{\frac{1}{2},0}|^{2}\\ \notag
&&+|H_{-\frac{1}{2},0}|^{2}+\frac{m_{l}^{2}}{2q^{2}}\Big(3|H_{\frac{1}{2},t}|^{2}+3|H_{-\frac{1}{2},t}|^{2}+|H_{\frac{1}{2},0}|^{2}+|H_{-\frac{1}{2},0}|^{2}\Big)\Bigg] \end{eqnarray}
\begin{eqnarray}\label{eq:36}
\notag
&&\frac{d\Gamma_{T}}{dq^{2}}=\frac{G_{F}^{2}|V_{CKM}|^{2}}{192\pi^{3}}\frac{m_{\mathcal{B}_{2}^{*}}}{m_{\mathcal{B}_{1}}^{2}}\frac{(q^{2}-m_{l}^{2})^{2}}{q^{2}}
\sqrt{\omega^2-1}\Bigg[|H_{\frac{1}{2},1}|^{2}\\ \notag
&&+|H_{-\frac{1}{2},-1}|^{2}+|H_{\frac{3}{2},1}|^{2}+|H_{-\frac{3}{2},-1}|^{2}+\frac{m_{l}^{2}}{2q^{2}}\Big(|H_{\frac{1}{2},1}|^{2}+|H_{-\frac{1}{2},-1}|^{2}\\
&&+|H_{\frac{3}{2},1}|^{2}+|H_{-\frac{3}{2},-1}|^{2}\Big)\Bigg]
\end{eqnarray}
The dependence of differential decay widths on $q^{2}$ are explicitly shown in Figs. \ref{decay1} and \ref{decay2}.
By integrating out the square momentum $q^{2}$, the decay widths can be written as,
\begin{eqnarray}\label{eq:37}
\Gamma=\int\limits_{m_{l}^{2}}^{(m_{\mathcal{B}_{1}}-m_{\mathcal{B}_{2}^{*}})^{2}}\Bigg\{\frac{d\Gamma_{L}}{dq^{2}}+\frac{d\Gamma_{T}}{dq^{2}}\Bigg\}dq^{2}
\end{eqnarray}

The integrated partial decay widths, ratios of $\Gamma_{L}/\Gamma_{T}$ and the corresponding branching fractions are calculated and the results are all listed in Tab. \ref{DW}. For comparison, the results in Ref. \cite{Hu:2020mxk} are listed in the last two columns in the table. In Ref. \cite{Gutsche:2019iac}, the authors used covariant confined quark model to analyze the decay processes of $\Xi_{cc}^{++}\rightarrow \Xi_{c}^{*\prime+}l^{+}\nu_{l}$, $\Omega_{cc}^{+}\rightarrow \Omega_{c}^{*0}l^{+}\nu_{l}$, where the decay widths are predicted to be $2.2\times10^{-14}$ and $4.0\times10^{-14}$ GeV. It can be seen that the predicted widths in present work are slightly lower than those obtained by quark model Refs. \cite{Hu:2020mxk,Gutsche:2019iac}. In addition, the predicted results in present work show that the decay channels $\Xi_{cc}^{++}\rightarrow\Xi_c^{\prime*+}l^{+}\nu_{l}$ and $\Omega_{cc}^{+}\rightarrow\Omega_c^{*0}l^{+}\nu_{l}$ have much larger branching fraction than the other decay modes.

In the flavor SU(3) symmetry, there exist the following relations for the semileptonic decay processes,
\begin{eqnarray}\label{eq:38}
\notag
&\Gamma\left(\Xi_{cc}^{++}\rightarrow\Sigma_{c}^{*+}l^{+}\nu\right)=\Gamma\left(\Omega_{cc}^{+}\rightarrow\Xi_{c}^{\prime*0}l^{+}\nu\right) \\
&\Gamma\left(\Xi_{cc}^{++}\rightarrow\Xi_{c}^{\prime*+}l^{+}\nu\right)=\frac{1}{2}\Gamma\left(\Omega_{cc}^{+}\rightarrow\Omega_{c}^{*0}l^{+}\nu\right)
\end{eqnarray}
The predictions in present work show that these relations are not strictly satisfied (See Tab. \ref{DW}), which reflects the SU(3) symmetry breaking effecting. The reason can be ascribed that the masses of $u$ and $d$ quark is taken to be zero in this work, however the mass of strange quark is not neglected.
\begin{table}[htbp]
\begin{ruledtabular}\caption{Results of semi-leptonic decay, where $\Gamma$ is in unit of $10^{-14}$ GeV and the lifetimes of $\Xi_{cc}$ and $\Omega_{cc}$ are taken to be 256 and 206 fs, respectively}
\label{DW}
\begin{tabular}{c| c c c |c c}
\multirow{2}{*}{Modes}& \multicolumn{3}{c|}{This work}& \multicolumn{2}{c}{Ref. \cite{Hu:2020mxk}} \\ \cline{2-6}
     &$\Gamma$&$\mathcal{B(\text{\textperthousand})}$&$\Gamma_{L}/\Gamma_{T}$&  $\Gamma$  & $\Gamma_{L}/\Gamma_{T}$ \\ \cline{2-6} \hline
$\Xi_{cc}^{++}\rightarrow\Sigma_c^{*+}e^{+}\nu_{e}$&$0.096^{+0.02}_{-0.01}$& $0.37^{+0.07}_{-0.05}$ & $1.65^{+0.12}_{-0.08}$ & 0.14& 0.92\\
$\Xi_{cc}^{++}\rightarrow\Sigma_c^{*+}\mu^{+}\nu_{\mu}$&$0.091^{+0.02}_{-0.01}$& $0.35^{+0.06}_{-0.05}$ & $1.62^{+0.12}_{-0.08}$ & 0.14&0.92\\
$\Xi_{cc}^{++}\rightarrow\Xi_c^{\prime*+}e^{+}\nu_{e}$&$1.01^{+0.24}_{-0.18}$& $3.93^{+0.93}_{-0.72}$ & $1.59^{+0.13}_{-0.10}$ & 1.74&1.08\\
$\Xi_{cc}^{++}\rightarrow\Xi_c^{\prime*+}\mu^{+}\nu_{\mu}$&$0.95^{+0.23}_{-0.17}$& $3.68^{+0.88}_{-0.68}$& $1.55^{+0.12}_{-0.10}$& 1.74&1.08\\
$\Omega_{cc}^{+}\rightarrow\Xi_c^{\prime*0}e^{+}\nu_{e}$&$0.085^{+0.02}_{-0.01}$&$0.27^{+0.07}_{-0.05}$ & $1.60^{+0.17}_{-0.11}$& 0.14&0.93\\
$\Omega_{cc}^{+}\rightarrow\Xi_c^{\prime*0}\mu^{+}\nu_{\mu}$&$0.081^{+0.02}_{-0.01}$&$0.25^{+0.06}_{-0.04}$&$1.57^{+0.17}_{-0.11}$& 0.14&0.93\\
$\Omega_{cc}^{+}\rightarrow\Omega_c^{*0}e^{+}\nu_{e}$&$1.95^{+0.53}_{-0.38}$&$4.82^{+0.45}_{-0.35}$&$1.54^{+0.14}_{-0.11}$& 3.45&1.07\\
$\Omega_{cc}^{+}\rightarrow\Omega_c^{*0}\mu^{+}\nu_{\mu}$&$1.83^{+0.50}_{-0.36}$&$4.69^{+0.44}_{-0.35}$&$1.50^{+0.14}_{-0.11}$& 3.45&1.07
		\end{tabular}
	\end{ruledtabular}
\end{table}

\section{Conclusions}\label{sec5}
In the present work, we firstly analyze the form factors of transition processes $\Xi_{cc}^{++}\rightarrow\Sigma_{c}^{*+}$, $\Xi_{cc}^{++}\rightarrow\Xi_{c}^{\prime*+}$, $\Omega_{cc}^{+}\rightarrow\Xi_{c}^{\prime*0}$ and $\Omega_{cc}^{++}\rightarrow\Omega_{c}^{*0}$ in the framework of three-point QCDSR. Considering the spin-parity of the initial and final baryons, these transition processes can be expressed as $\frac{1}{2}^{+}\rightarrow\frac{3}{2}^{+}$. In the phenomenological side, if all couplings of interpolating currents to hadronic states are considered, there are many redundant contributions to the results. For example, the states with $J^{P}=\frac{1}{2}^{-}$ will lead to contaminations to the initial baryons, and $J^{P}=\frac{1}{2}^{\pm}$, $\frac{3}{2}^{-}$ states will interfere the final baryons. By eliminating the dirac structures $\gamma_{\mu}$ and $p_{\mu}^{\prime}$ and solving 16 linear equations, we successfully extract the form factors for the process $\frac{1}{2}^{+}\rightarrow\frac{3}{2}^{+}$. As for the OPE in the QCD side, the perturbative part, vacuum condensate terms of $\langle{\bar qq}\rangle$, $\langle g_{s}^{2}GG\rangle$, $\langle \bar q g_{s}\sigma Gq\rangle$ and $g_{s}^{2}\langle{\bar qq}\rangle^{2}$ are considered. After the $Q^{2}$-dependence of the form factors are calculated, the numerical results are fitted into analytical function and then extrapolated into time-like region ($Q^2<0$). Using the predicted form factors, we analyze the decay properties of $\Xi_{cc}^{++}\rightarrow \Sigma_{c}^{*+}l^{+}\nu_{l}$, $\Xi_{cc}^{++}\rightarrow \Xi_{c}^{\prime*+}l^{+}\nu_{l}$, $\Omega_{cc}^{+}\rightarrow\Xi_{c}^{\prime*0}l^{+}\nu_{l}$ and $\Omega_{cc}^{+}\rightarrow \Omega_{c}^{*0}l^{+}\nu_{l}$. Although our predicted values of form factors at zero are not consistent with those of other literatures, the difference becomes smaller after the numerical results are extrapolated into time-like region. Thus, the final decay widths predicted in present work are comparable with those of other collaborations.

More theoretical efforts in analyzing the weak decays of singly- and doubly-heavy baryons are still needed. Because the form factors that we have calculated in the present work are related to heavy to light transitions at quark level, the heavy to heavy transition will be studied in the future. In addition. the form factors studied now are defined by vector and axial-vector currents, while the tensor form factors are also important parameters in some other decay processes such as the radiative and the dilepton decay modes.

\section*{Acknowledgements}
This project is supported by National Natural Science Foundation under the Grant No. 12575083, and Natural Science Foundation of HeBei Province under the Grant No. A2024202008.

\begin{widetext}
\appendix
\section{Form factor of $\mathcal{B}_{1}\rightarrow \mathcal{B}_{2}^{*}$}\label{Sec:AppA}
\begin{eqnarray}\label{eq:A1}
\notag
\left\langle \mathcal{B}_{2}^{*+}\left(p^{\prime}\right)|J^{V-A}_{\nu}|\mathcal{B}_{1}^{+}\left(p\right) \right\rangle&&=\overline{u}_{\alpha}^{\mathcal{B}_{2}^{*+}}\left(p^{\prime}\right)\gamma_{5}\left[\gamma_{\nu}\frac{p_{\alpha}}{m_{\mathcal{B}_{1}^{+}}}F_{1}^{++}\left(q^{2}\right)
+\frac{p_{\alpha}p_{\nu}}{m_{\mathcal{B}_{1}^{+}}^{2}}F_{2}^{++}\left(q^{2}\right)+\frac{p_{\alpha}p^{\prime}_{\nu}}{m_{\mathcal{B}_{1}^{+}}m_{\mathcal{B}_{2}^{*+}}}F_{3}^{++}\left(q^{2}\right)+g_{\alpha\nu}F_{4}^{++}\left(q^{2}\right)\right]u^{\mathcal{B}_{1}^{+}}\left(p\right) \\ \notag
&&-\overline{u}_{\alpha}^{\mathcal{B}_{2}^{*+}}\left(p^{\prime}\right)\left[\gamma_{\nu}\frac{p_{\alpha}}{m_{\mathcal{B}_{1}^{+}}}G_{1}^{++}\left(q^{2}\right)
+\frac{p_{\alpha}p_{\nu}}{m_{\mathcal{B}_{1}^{+}}^{2}}G_{2}^{++}\left(q^{2}\right)+\frac{p_{\alpha}p^{\prime}_{\nu}}{m_{\mathcal{B}_{1}^{+}}m_{\mathcal{B}_{2}^{*+}}}G_{3}^{++}\left(q^{2}\right)+g_{\alpha\nu}G_{4}^{++}\left(q^{2}\right)\right]u^{\mathcal{B}_{1}^{+}}\left(p\right)\\
\notag
\left\langle \mathcal{B}_{2}^{*-}\left(p^{\prime}\right)|J^{V-A}_{\nu}|\mathcal{B}_{1}^{+}\left(p\right)  \right\rangle&&=\overline{u}_{\alpha}^{\mathcal{B}_{2}^{*-}}\left(p^{\prime}\right)\left[\gamma_{\nu}\frac{p_{\alpha}}{m_{\mathcal{B}_{1}^{+}}}F_{1}^{+-}\left(q^{2}\right)
+\frac{p_{\alpha}p_{\nu}}{m_{\mathcal{B}_{1}^{+}}^{2}}F_{2}^{+-}\left(q^{2}\right)+\frac{p_{\alpha}p^{\prime}_{\nu}}{m_{\mathcal{B}_{1}^{+}}m_{\mathcal{B}_{2}^{*-}}}F_{3}^{+-}\left(q^{2}\right)+g_{\alpha\nu}F_{4}^{+-}\left(q^{2}\right)\right]u^{\mathcal{B}_{1}^{+}}\left(p\right) \\ \notag
&&-\overline{u}_{\alpha}^{\mathcal{B}_{2}^{*-}}\left(p^{\prime}\right)\gamma_{5}\left[\gamma_{\nu}\frac{p_{\alpha}}{m_{\mathcal{B}_{1}^{+}}}G_{1}^{+-}\left(q^{2}\right)
+\frac{p_{\alpha}p_{\nu}}{m_{\mathcal{B}_{1}^{+}}^{2}}G_{2}^{+-}\left(q^{2}\right)+\frac{p_{\alpha}p^{\prime}_{\nu}}{m_{\mathcal{B}_{1}^{+}}m_{\mathcal{B}_{2}^{*-}}}G_{3}^{+-}\left(q^{2}\right)+g_{\alpha\nu}G_{4}^{+-}\left(q^{2}\right)\right]u^{\mathcal{B}_{1}^{+}}\left(p\right)\\
\notag
\left\langle \mathcal{B}_{2}^{*+}\left(p^{\prime}\right)|J^{V-A}_{\nu}|\mathcal{B}_{1}^{-}\left(p\right)  \right\rangle&&=\overline{u}_{\alpha}^{\mathcal{B}_{2}^{*+}}\left(p^{\prime}\right)\left[\gamma_{\nu}\frac{p_{\alpha}}{m_{\mathcal{B}_{1}^{-}}}F_{1}^{-+}\left(q^{2}\right)
+\frac{p_{\alpha}p_{\nu}}{m_{\mathcal{B}_{1}^{-}}^{2}}F_{2}^{-+}\left(q^{2}\right)+\frac{p_{\alpha}p^{\prime}_{\nu}}{m_{\mathcal{B}_{1}^{-}}m_{\mathcal{B}_{2}^{*+}}}F_{3}^{-+}\left(q^{2}\right)+g_{\alpha\nu}F_{4}^{-+}\left(q^{2}\right)\right]u^{\mathcal{B}_{1}^{-}}\left(p\right) \\ \notag
&&-\overline{u}_{\alpha}^{\mathcal{B}_{2}^{*+}}\left(p^{\prime}\right)\gamma_{5}\left[\gamma_{\nu}\frac{p_{\alpha}}{m_{\mathcal{B}_{1}^{-}}}G_{1}^{-+}\left(q^{2}\right)
+\frac{p_{\alpha}p_{\nu}}{m_{\mathcal{B}_{1}^{-}}^{2}}G_{2}^{-+}\left(q^{2}\right)+\frac{p_{\alpha}p^{\prime}_{\nu}}{m_{\mathcal{B}_{1}^{-}}m_{\mathcal{B}_{2}^{*+}}}G_{3}^{-+}\left(q^{2}\right)+g_{\alpha\nu}G_{4}^{-+}\left(q^{2}\right)\right]u^{\mathcal{B}_{1}^{-}}\left(p\right)\\
\notag
\left\langle \mathcal{B}_{2}^{*-}\left(p^{\prime}\right)|J^{V-A}_{\nu}|\mathcal{B}_{1}^{-}\left(p\right) \right\rangle&&=\overline{u}_{\alpha}^{\mathcal{B}_{2}^{*-}}\left(p^{\prime}\right)\gamma_{5}\left[\gamma_{\nu}\frac{p_{\alpha}}{m_{\mathcal{B}_{1}^{-}}}F_{1}^{--}\left(q^{2}\right)
+\frac{p_{\alpha}p_{\nu}}{m_{\mathcal{B}_{1}^{-}}^{2}}F_{2}^{--}\left(q^{2}\right)+\frac{p_{\alpha}p^{\prime}_{\nu}}{m_{\mathcal{B}_{1}^{-}}m_{\mathcal{B}_{2}^{*-}}}F_{3}^{--}\left(q^{2}\right)+g_{\alpha\nu}F_{4}^{--}\left(q^{2}\right)\right]u^{\mathcal{B}_{1}^{-}}\left(p\right) \\ \notag
&&-\overline{u}_{\alpha}^{\mathcal{B}_{2}^{*-}}\left(p^{\prime}\right)\left[\gamma_{\nu}\frac{p_{\alpha}}{m_{\mathcal{B}_{1}^{-}}}G_{1}^{--}\left(q^{2}\right)
+\frac{p_{\alpha}p_{\nu}}{m_{\mathcal{B}_{1}^{-}}^{2}}G_{2}^{--}\left(q^{2}\right)+\frac{p_{\alpha}p^{\prime}_{\nu}}{m_{\mathcal{B}_{1}^{-}}m_{\mathcal{B}_{2}^{*-}}}G_{3}^{--}\left(q^{2}\right)+g_{\alpha\nu}G_{4}^{--}\left(q^{2}\right)\right]u^{\mathcal{B}_{1}^{-}}\left(p\right)\\
\notag
\left\langle \mathcal{B}_{2}^{+}\left(p^{\prime}\right)|J^{V-A}_{\nu}|\mathcal{B}_{1}^{+}\left(p\right) \right\rangle&&=\overline{u}^{\mathcal{B}_{2}^{+}}\left(p^{\prime}\right)\left[\frac{p_{\nu}}{m_{\mathcal{B}_{1}^{+}}}F_{1}^{\prime++}\left(q^{2}\right)
+\frac{p^{\prime}_{\nu}}{m_{\mathcal{B}_{2}^{+}}}F_{2}^{\prime++}\left(q^{2}\right)+\gamma_{\nu}F_{3}^{\prime++}\left(q^{2}\right)\right]u^{\mathcal{B}_{1}^{+}}\left(p\right)
\\ \notag
&&-\overline{u}^{\mathcal{B}_{2}^{+}}\left(p^{\prime}\right)\left[\frac{p_{\nu}}{m_{\mathcal{B}_{1}^{+}}}G_{1}^{\prime++}\left(q^{2}\right)
+\frac{p^{\prime}_{\nu}}{m_{\mathcal{B}_{2}^{+}}}G_{2}^{\prime++}\left(q^{2}\right)+\gamma_{\nu}G_{3}^{\prime++}\left(q^{2}\right)\right]\gamma_{5}u^{\mathcal{B}_{1}^{+}}\left(p\right)
\\ \notag
\left\langle \mathcal{B}_{2}^{-}\left(p^{\prime}\right)|J^{V-A}_{\nu}|\mathcal{B}_{1}^{+}\left(p\right) \right\rangle&&=\overline{u}^{\mathcal{B}_{2}^{-}}\left(p^{\prime}\right)\gamma_{5}\left[\frac{p_{\nu}}{m_{\mathcal{B}_{1}^{+}}}F_{1}^{\prime+-}\left(q^{2}\right)
+\frac{p^{\prime}_{\nu}}{m_{\mathcal{B}_{2}^{-}}}F_{2}^{\prime+-}\left(q^{2}\right)+\gamma_{\nu}F_{3}^{\prime+-}\left(q^{2}\right)\right]u^{\mathcal{B}_{1}^{+}}\left(p\right)
\\ \notag
&&-\overline{u}^{\mathcal{B}_{2}^{-}}\left(p^{\prime}\right)\gamma_{5}\left[\frac{p_{\nu}}{m_{\mathcal{B}_{1}^{+}}}G_{1}^{\prime+-}\left(q^{2}\right)
+\frac{p^{\prime}_{\nu}}{m_{\mathcal{B}_{2}^{-}}}G_{2}^{\prime+-}\left(q^{2}\right)+\gamma_{\nu}G_{3}^{\prime+-}\left(q^{2}\right)\right]\gamma_{5}u^{\mathcal{B}_{1}^{+}}\left(p\right)
\\ \notag
\left\langle \mathcal{B}_{2}^{+}\left(p^{\prime}\right)|J^{V-A}_{\nu}|\mathcal{B}_{1}^{-}\left(p\right) \right\rangle&&=\overline{u}^{\mathcal{B}_{2}^{+}}\left(p^{\prime}\right)\left[\frac{p_{\nu}}{m_{\mathcal{B}_{1}^{-}}}F_{1}^{\prime-+}\left(q^{2}\right)
+\frac{p^{\prime}_{\nu}}{m_{\mathcal{B}_{2}^{+}}}F_{2}^{\prime-+}\left(q^{2}\right)+\gamma_{\nu}F_{3}^{\prime-+}\left(q^{2}\right)\right]\gamma_{5}u^{\mathcal{B}_{1}^{-}}\left(p\right)
\\ \notag
&&-\overline{u}^{\mathcal{B}_{2}^{+}}\left(p^{\prime}\right)\left[\frac{p_{\nu}}{m_{\mathcal{B}_{1}^{-}}}G_{1}^{\prime-+}\left(q^{2}\right)
+\frac{p^{\prime}_{\nu}}{m_{\mathcal{B}_{2}^{+}}}G_{2}^{\prime-+}\left(q^{2}\right)+\gamma_{\nu}G_{3}^{\prime-+}\left(q^{2}\right)\right]\gamma_{5}\gamma_{5}u^{\mathcal{B}_{1}^{-}}\left(p\right)
\\ \notag
\left\langle \mathcal{B}_{2}^{-}\left(p^{\prime}\right)|J^{V-A}_{\nu}|\mathcal{B}_{1}^{-}\left(p\right) \right\rangle&&=\overline{u}^{\mathcal{B}_{2}^{-}}\left(p^{\prime}\right)\gamma_{5}\left[\frac{p_{\nu}}{m_{\mathcal{B}_{1}^{-}}}F_{1}^{\prime--}\left(q^{2}\right)
+\frac{p^{\prime}_{\nu}}{m_{\mathcal{B}_{2}^{-}}}F_{2}^{\prime--}\left(q^{2}\right)+\gamma_{\nu}F_{3}^{\prime--}\left(q^{2}\right)\right]\gamma_{5}u^{\mathcal{B}_{1}^{-}}\left(p\right)
\\
&&-\overline{u}^{\mathcal{B}_{2}^{-}}\left(p^{\prime}\right)\gamma_{5}\left[\frac{p_{\nu}}{m_{\mathcal{B}_{1}^{-}}}G_{1}^{\prime--}\left(q^{2}\right)
+\frac{p^{\prime}_{\nu}}{m_{\mathcal{B}_{2}^{-}}}G_{2}^{\prime--}\left(q^{2}\right)+\gamma_{\nu}G_{3}^{\prime--}\left(q^{2}\right)\right]\gamma_{5}\gamma_{5}u^{\mathcal{B}_{1}^{-}}\left(p\right)
\end{eqnarray}
\begin{eqnarray}\label{eq:A2}
\notag
&&\Pi _{\mu\nu}^{\mathrm{phy}}\left(p,p^{\prime}\right)= \\ \notag
&&\frac{\lambda_{\mathcal{B}_{2}^{*+}}\lambda_{\mathcal{B}_{1}^{+}}u_{\mu}^{\mathcal{B}_{2}^{*+}}\left(p^{\prime},s^{\prime}\right)\overline{u}_{\alpha}^{\mathcal{B}_{2}^{*+}}\left(p^{\prime},s^{\prime}\right)\gamma_{5}\left[\frac{p_{\alpha}\gamma_{\nu}}{m_{\mathcal{B}_{1}^{+}}}F_{1}^{++}\left(q^{2}\right)
+\frac{p_{\alpha}p_{\nu}}{m_{\mathcal{B}_{1}^{+}}^{2}}F_{2}^{++}\left(q^{2}\right)+\frac{p_{\alpha}p^{\prime}_{\nu}}{m_{\mathcal{B}_{1}^{+}}m_{\mathcal{B}_{2}^{*+}}}F_{3}^{++}\left(q^{2}\right)+g_{\alpha\nu}F_{4}^{++}\left(q^{2}\right)\right]u^{B^{+}_{1}}\left(p,s\right)\overline{u}^{B^{+}_{1}}\left(p,s\right)}{{\left(p{'^2} - m_{{\mathcal{B}^{*+}_{2}}}^2\right)\left({p^2} - m_{\mathcal{B}_{1}^{+}}^2\right)}} \\ \notag
&&-\frac{\lambda_{\mathcal{B}_{2}^{*+}}\lambda_{\mathcal{B}_{1}^{+}}u_{\mu}^{\mathcal{B}_{2}^{*+}}\left(p^{\prime},s^{\prime}\right)\overline{u}_{\alpha}^{\mathcal{B}_{2}^{*+}}\left(p^{\prime},s^{\prime}\right)\left[\frac{p_{\alpha}\gamma_{\nu}}{m_{\mathcal{B}_{1}^{+}}}G_{1}^{++}\left(q^{2}\right)
+\frac{p_{\alpha}p_{\nu}}{m_{\mathcal{B}_{1}^{+}}^{2}}G_{2}^{++}\left(q^{2}\right)+\frac{p_{\alpha}p^{\prime}_{\nu}}{m_{\mathcal{B}_{1}^{+}}m_{\mathcal{B}_{2}^{*+}}}G_{3}^{++}\left(q^{2}\right)+g_{\alpha\nu}G_{4}^{++}\left(q^{2}\right)\right]u^{B^{+}_{1}}\left(p,s\right)\overline{u}^{B^{+}_{1}}\left(p,s\right)}{{\left(p{'^2} - m_{{\mathcal{B}^{*+}_{2}}}^2\right)\left({p^2} - m_{\mathcal{B}_{1}^{+}}^2\right)}} \\ \notag
&&+\frac{\lambda_{\mathcal{B}_{2}^{*-}}\lambda_{\mathcal{B}_{1}^{+}}u_{\mu}^{\mathcal{B}_{2}^{*-}}\left(p^{\prime},s^{\prime}\right)\overline{u}_{\alpha}^{\mathcal{B}_{2}^{*-}}\left(p^{\prime},s^{\prime}\right)\left[\frac{p_{\alpha}\gamma_{\nu}}{m_{\mathcal{B}_{1}^{+}}}F_{1}^{+-}\left(q^{2}\right)
+\frac{p_{\alpha}p_{\nu}}{m_{\mathcal{B}_{1}^{+}}^{2}}F_{2}^{+-}\left(q^{2}\right)+\frac{p_{\alpha}p^{\prime}_{\nu}}{m_{\mathcal{B}_{1}^{+}}m_{\mathcal{B}_{2}^{*-}}}F_{3}^{+-}\left(q^{2}\right)+g_{\alpha\nu}F_{4}^{+-}\left(q^{2}\right)\right]u^{B^{+}_{1}}\left(p,s\right)\overline{u}^{B^{+}_{1}}\left(p,s\right)}{{\left(p{'^2} - m_{{\mathcal{B}^{*-}_{2}}}^2\right)\left({p^2} - m_{\mathcal{B}_{1}^{+}}^2\right)}} \\ \notag
&&-\frac{\lambda_{\mathcal{B}_{2}^{*-}}\lambda_{\mathcal{B}_{1}^{+}}u_{\mu}^{\mathcal{B}_{2}^{*-}}\left(p^{\prime},s^{\prime}\right)\overline{u}_{\alpha}^{\mathcal{B}_{2}^{*-}}\left(p^{\prime},s^{\prime}\right)\gamma_{5}\left[\frac{p_{\alpha}\gamma_{\nu}}{m_{\mathcal{B}_{1}^{+}}}G_{1}^{+-}\left(q^{2}\right)
+\frac{p_{\alpha}p_{\nu}}{m_{\mathcal{B}_{1}^{+}}^{2}}G_{2}^{+-}\left(q^{2}\right)+\frac{p_{\alpha}p^{\prime}_{\nu}}{m_{\mathcal{B}_{1}^{+}}m_{\mathcal{B}_{2}^{*-}}}G_{3}^{+-}\left(q^{2}\right)+g_{\alpha\nu}G_{4}^{+-}\left(q^{2}\right)\right]u^{B^{+}_{1}}\left(p,s\right)\overline{u}^{B^{+}_{1}}\left(p,s\right)}{{\left(p{'^2} - m_{{\mathcal{B}^{*-}_{2}}}^2\right)\left({p^2} - m_{\mathcal{B}_{1}^{+}}^2\right)}} \\ \notag
&&+\frac{\lambda_{\mathcal{B}_{2}^{*+}}\lambda_{\mathcal{B}_{1}^{-}}u_{\mu}^{\mathcal{B}_{2}^{*+}}\left(p^{\prime},s^{\prime}\right)\overline{u}_{\alpha}^{\mathcal{B}_{2}^{*+}}\left(p^{\prime},s^{\prime}\right)\left[\frac{p_{\alpha}\gamma_{\nu}}{m_{\mathcal{B}_{1}^{-}}}F_{1}^{-+}\left(q^{2}\right)
+\frac{p_{\alpha}p_{\nu}}{m_{\mathcal{B}_{1}^{-}}^{2}}F_{2}^{-+}\left(q^{2}\right)+\frac{p_{\alpha}p^{\prime}_{\nu}}{m_{\mathcal{B}_{1}^{-}}m_{\mathcal{B}_{2}^{*+}}}F_{3}^{-+}\left(q^{2}\right)+g_{\alpha\nu}F_{4}^{-+}\left(q^{2}\right)\right]u^{B^{-}_{1}}\left(p,s\right)\overline{u}^{B^{-}_{1}}\left(p,s\right)}{{\left(p{'^2} - m_{{\mathcal{B}^{*+}_{2}}}^2\right)\left({p^2} - m_{\mathcal{B}_{1}^{-}}^2\right)}} \\ \notag
&&-\frac{\lambda_{\mathcal{B}_{2}^{*+}}\lambda_{\mathcal{B}_{1}^{-}}u_{\mu}^{\mathcal{B}_{2}^{*+}}\left(p^{\prime},s^{\prime}\right)\overline{u}_{\alpha}^{\mathcal{B}_{2}^{*+}}\left(p^{\prime},s^{\prime}\right)\gamma_{5}\left[\frac{p_{\alpha}\gamma_{\nu}}{m_{\mathcal{B}_{1}^{-}}}G_{1}^{-+}\left(q^{2}\right)
+\frac{p_{\alpha}p_{\nu}}{m_{\mathcal{B}_{1}^{-}}^{2}}G_{2}^{-+}\left(q^{2}\right)+\frac{p_{\alpha}p^{\prime}_{\nu}}{m_{\mathcal{B}_{1}^{-}}m_{\mathcal{B}_{2}^{*+}}}G_{3}^{-+}\left(q^{2}\right)+g_{\alpha\nu}G_{4}^{-+}\left(q^{2}\right)\right]u^{B^{-}_{1}}\left(p,s\right)\overline{u}^{B^{-}_{1}}\left(p,s\right)}{{\left(p{'^2} - m_{{\mathcal{B}^{*+}_{2}}}^2\right)\left({p^2} - m_{\mathcal{B}_{1}^{-}}^2\right)}} \\ \notag
&&+\frac{\lambda_{\mathcal{B}_{2}^{*-}}\lambda_{\mathcal{B}_{1}^{-}}u_{\mu}^{\mathcal{B}_{2}^{*-}}\left(p^{\prime},s^{\prime}\right)\overline{u}_{\alpha}^{\mathcal{B}_{2}^{*-}}\left(p^{\prime},s^{\prime}\right)\gamma_{5}\left[\frac{p_{\alpha}\gamma_{\nu}}{m_{\mathcal{B}_{1}^{-}}}F_{1}^{--}\left(q^{2}\right)
+\frac{p_{\alpha}p_{\nu}}{m_{\mathcal{B}_{1}^{-}}^{2}}F_{2}^{--}\left(q^{2}\right)+\frac{p_{\alpha}p^{\prime}_{\nu}}{m_{\mathcal{B}_{1}^{-}}m_{\mathcal{B}_{2}^{*-}}}F_{3}^{--}\left(q^{2}\right)+g_{\alpha\nu}F_{4}^{--}\left(q^{2}\right)\right]u^{B^{-}_{1}}\left(p,s\right)\overline{u}^{B^{-}_{1}}\left(p,s\right)}{{\left(p{'^2} - m_{{\mathcal{B}^{*-}_{2}}}^2\right)\left({p^2} - m_{\mathcal{B}_{1}^{-}}^2\right)}} \\ \notag
&&-\frac{\lambda_{\mathcal{B}_{2}^{*-}}\lambda_{\mathcal{B}_{1}^{-}}u_{\mu}^{\mathcal{B}_{2}^{*-}}\left(p^{\prime},s^{\prime}\right)\overline{u}_{\alpha}^{\mathcal{B}_{2}^{*-}}\left(p^{\prime},s^{\prime}\right)\left[\frac{p_{\alpha}\gamma_{\nu}}{m_{\mathcal{B}_{1}^{-}}}G_{1}^{--}\left(q^{2}\right)
+\frac{p_{\alpha}p_{\nu}}{m_{\mathcal{B}_{1}^{-}}^{2}}G_{2}^{--}\left(q^{2}\right)+\frac{p_{\alpha}p^{\prime}_{\nu}}{m_{\mathcal{B}_{1}^{-}}m_{\mathcal{B}_{2}^{*-}}}G_{3}^{--}\left(q^{2}\right)+g_{\alpha\nu}G_{4}^{--}\left(q^{2}\right)\right]u^{B^{-}_{1}}\left(p,s\right)\overline{u}^{B^{-}_{1}}\left(p,s\right)}{{\left(p{'^2} - m_{{\mathcal{B}^{*-}_{2}}}^2\right)\left({p^2} - m_{\mathcal{B}_{1}^{-}}^2\right)}}\\
&&+\cdots
\end{eqnarray}	
\begin{eqnarray}\label{eq:A3}
\notag
F_{1}^{++}(Q^2) &&=
\frac{m_{\mathcal{B}_{1}^{+}}e^{\frac{m_{\mathcal{B}_{1}^{+}}^{2}}{M_{1}^{2}}+\frac{m_{\mathcal{B}_{2}^{*+}}^{2}}{M_{2}^{2}}}}
{\left(m_{\mathcal{B}_{2}^{*+}}+m_{\mathcal{B}_{2}^{*-}}\right)\left(m_{\mathcal{B}_{1}^{+}}+m_{\mathcal{B}_{1}^{-}}\right)\lambda_{\mathcal{B}_{2}^{*+}}\lambda_{\mathcal{B}_{1}^{+}}}
\int^{s_{0}}_{s_{min}}ds\int^{u_{0}}_{u_{min}}du\Big\{\left[Q^{2}-m_{\mathcal{B}_{2}^{*-}}m_{\mathcal{B}_{1}^{-}}+m_{\mathcal{B}_{1}^{+}}m_{\mathcal{B}_{1}^{-}}+m_{\mathcal{B}_{2}^{*+}}m_{\mathcal{B}_{2}^{*-}}\right]\rho_{4}^{\mathrm{QCD-V}}\left(s,u,Q^2\right)
\\ \notag
&&+\rho_{1}^{\mathrm{QCD-V}}\left(s,u,Q^2\right)+\left[m_{\mathcal{B}_{2}^{*+}}-m_{\mathcal{B}_{2}^{*-}}-m_{\mathcal{B}_{1}^{-}}\right]\rho_{2}^{\mathrm{QCD-V}}\left(s,u,Q^2\right)
+\left[m_{\mathcal{B}_{1}^{+}}-m_{\mathcal{B}_{1}^{-}}-m_{\mathcal{B}_{2}^{*-}}\right]\rho_{3}^{\mathrm{QCD-V}}\left(s,u,Q^2\right)\Big\} \\ \notag
F_{2}^{++}(Q^2) &&=
-\frac{m_{\mathcal{B}_{1}^{+}}^{2}e^{\frac{m_{\mathcal{B}_{1}^{+}}^{2}}{M_{1}^{2}}+\frac{m_{\mathcal{B}_{2}^{*+}}^{2}}{M_{2}^{2}}}}
{\left(m_{\mathcal{B}_{2}^{*+}}+m_{\mathcal{B}_{2}^{*-}}\right)\left(m_{\mathcal{B}_{1}^{+}}+m_{\mathcal{B}_{1}^{-}}\right)\lambda_{\mathcal{B}_{2}^{*+}}\lambda_{\mathcal{B}_{1}^{+}}}
\int^{s_{0}}_{s_{min}}ds\int^{u_{0}}_{u_{min}}du\Big\{\left[Q^{2}+m_{\mathcal{B}_{2}^{*-}}m_{\mathcal{B}_{1}^{-}}+m_{\mathcal{B}_{1}^{+}}m_{\mathcal{B}_{1}^{-}}
+m_{\mathcal{B}_{2}^{*+}}m_{\mathcal{B}_{2}^{*-}}\right]\rho_{12}^{\mathrm{QCD-V}}\left(s,u,Q^2\right)
\\ \notag
&&+\rho_{9}^{\mathrm{QCD-V}}\left(s,u,Q^2\right)+\left[m_{\mathcal{B}_{2}^{*-}}+m_{\mathcal{B}_{1}^{+}}-m_{\mathcal{B}_{1}^{-}}\right]\rho_{11}^{\mathrm{QCD-V}}\left(s,u,Q^2\right)
+\left[m_{\mathcal{B}_{2}^{*-}}-m_{\mathcal{B}_{2}^{*+}}-m_{\mathcal{B}_{1}^{-}}\right]\rho_{10}^{\mathrm{QCD-V}}\left(s,u,Q^2\right)\Big\} \\ \notag
F_{3}^{++}(Q^2) &&=
-\frac{m_{\mathcal{B}_{1}^{+}}m_{\mathcal{B}_{2}^{*+}}e^{\frac{m_{\mathcal{B}_{1}^{+}}^{2}}{M_{1}^{2}}+\frac{m_{\mathcal{B}_{2}^{*+}}^{2}}{M_{2}^{2}}}}
{\left(m_{\mathcal{B}_{2}^{*+}}+m_{\mathcal{B}_{2}^{*-}}\right)\left(m_{\mathcal{B}_{1}^{+}}+m_{\mathcal{B}_{1}^{-}}\right)\lambda_{\mathcal{B}_{2}^{*+}}\lambda_{\mathcal{B}_{1}^{+}}}
\int^{s_{0}}_{s_{min}}ds\int^{u_{0}}_{u_{min}}du\Big\{\left[Q^{2}+m_{\mathcal{B}_{2}^{*-}}m_{\mathcal{B}_{1}^{-}}+m_{\mathcal{B}_{1}^{+}}m_{\mathcal{B}_{1}^{-}}
+m_{\mathcal{B}_{2}^{*+}}m_{\mathcal{B}_{2}^{*-}}\right]\rho_{16}^{\mathrm{QCD-V}}\left(s,u,Q^2\right)
\\ \notag
&&+\rho_{13}^{\mathrm{QCD-V}}\left(s,u,Q^2\right)+\left[m_{\mathcal{B}_{2}^{*-}}-m_{\mathcal{B}_{1}^{-}}-m_{\mathcal{B}_{2}^{*+}}\right]\rho_{14}^{\mathrm{QCD-V}}\left(s,u,Q^2\right)
+\left[m_{\mathcal{B}_{2}^{*-}}+m_{\mathcal{B}_{1}^{+}}-m_{\mathcal{B}_{1}^{-}}\right]\rho_{15}^{\mathrm{QCD-V}}\left(s,u,Q^2\right) \\ \notag
&&+2\rho_{3}^{\mathrm{QCD-V}}\left(s,u,Q^2\right)+2\left(m_{\mathcal{B}_{1}^{-}}-m_{\mathcal{B}_{2}^{*-}}+m_{\mathcal{B}_{2}^{*+}}\right)\rho_{4}^{\mathrm{QCD-V}}\Big\} \\ \notag
F_{4}^{++}(Q^2) &&=
-\frac{e^{\frac{m_{\mathcal{B}_{1}^{+}}^{2}}{M_{1}^{2}}+\frac{m_{\mathcal{B}_{2}^{*+}}^{2}}{M_{2}^{2}}}}
{\left(m_{\mathcal{B}_{2}^{*+}}+m_{\mathcal{B}_{2}^{*-}}\right)\left(m_{\mathcal{B}_{1}^{+}}+m_{\mathcal{B}_{1}^{-}}\right)\lambda_{\mathcal{B}_{2}^{*+}}\lambda_{\mathcal{B}_{1}^{+}}}
\int^{s_{0}}_{s_{min}}ds\int^{u_{0}}_{u_{min}}du\Big\{\left[Q^{2}+m_{\mathcal{B}_{2}^{*+}}m_{\mathcal{B}_{2}^{*-}}+m_{\mathcal{B}_{2}^{*-}}m_{\mathcal{B}_{1}^{-}}+m_{\mathcal{B}_{1}^{+}}m_{\mathcal{B}_{1}^{-}}
\right]\rho_{8}^{\mathrm{QCD-V}}\left(s,u,Q^2\right) \\ \notag
&&+\rho_{5}^{\mathrm{QCD-V}}\left(s,u,Q^2\right)+\left[m_{\mathcal{B}_{2}^{*-}}-m_{\mathcal{B}_{1}^{-}}-m_{\mathcal{B}_{2}^{*+}}\right]\rho_{6}^{\mathrm{QCD-V}}\left(s,u,Q^2\right)
+\left[m_{\mathcal{B}_{2}^{*-}}+m_{\mathcal{B}_{1}^{+}}-m_{\mathcal{B}_{1}^{-}}\right]\rho_{7}^{\mathrm{QCD-V}}\left(s,u,Q^2\right) \Big\}  \\
\notag
G_{1}^{++}(Q^2) &&=
-\frac{m_{\mathcal{B}_{1}^{+}}e^{\frac{m_{\mathcal{B}_{1}^{+}}^{2}}{M_{1}^{2}}+\frac{m_{\mathcal{B}_{2}^{*+}}^{2}}{M_{2}^{2}}}}
{\left(m_{\mathcal{B}_{2}^{*+}}+m_{\mathcal{B}_{2}^{*-}}\right)\left(m_{\mathcal{B}_{1}^{+}}+m_{\mathcal{B}_{1}^{-}}\right)\lambda_{\mathcal{B}_{2}^{*+}}\lambda_{\mathcal{B}_{1}^{+}}}
\int^{s_{0}}_{s_{min}}ds\int^{u_{0}}_{u_{min}}du\Big\{\left[Q^{2}+m_{\mathcal{B}_{2}^{*-}}m_{\mathcal{B}_{1}^{-}}+m_{\mathcal{B}_{1}^{+}}m_{\mathcal{B}_{1}^{-}}+m_{\mathcal{B}_{2}^{*+}}m_{\mathcal{B}_{2}^{*-}}\right]\rho_{4}^{\mathrm{QCD-A}}\left(s,u,Q^2\right)
\\ \notag
&&+\rho_{1}^{\mathrm{QCD-A}}\left(s,u,Q^2\right)+\left[m_{\mathcal{B}_{2}^{*+}}-m_{\mathcal{B}_{2}^{*-}}+m_{\mathcal{B}_{1}^{-}}\right]\rho_{2}^{\mathrm{QCD-A}}\left(s,u,Q^2\right)
+\left[m_{\mathcal{B}_{1}^{-}}-m_{\mathcal{B}_{1}^{+}}-m_{\mathcal{B}_{2}^{*-}}\right]\rho_{3}^{\mathrm{QCD-A}}\left(s,u,Q^2\right)\Big\} \\ \notag
G_{2}^{++}(Q^2) &&=
-\frac{m_{\mathcal{B}_{1}^{+}}^{2}e^{\frac{m_{\mathcal{B}_{1}^{+}}^{2}}{M_{1}^{2}}+\frac{m_{\mathcal{B}_{2}^{*+}}^{2}}{M_{2}^{2}}}}
{\left(m_{\mathcal{B}_{2}^{*+}}+m_{\mathcal{B}_{2}^{*-}}\right)\left(m_{\mathcal{B}_{1}^{+}}+m_{\mathcal{B}_{1}^{-}}\right)\lambda_{\mathcal{B}_{2}^{*+}}\lambda_{\mathcal{B}_{1}^{+}}}
\int^{s_{0}}_{s_{min}}ds\int^{u_{0}}_{u_{min}}du\Big\{\left[Q^{2}+m_{\mathcal{B}_{2}^{*-}}m_{\mathcal{B}_{2}^{*+}}+m_{\mathcal{B}_{1}^{+}}m_{\mathcal{B}_{1}^{-}}
-m_{\mathcal{B}_{2}^{*-}}m_{\mathcal{B}_{1}^{-}}\right]\rho_{9}^{\mathrm{QCD-A}}\left(s,u,Q^2\right)
\\ \notag
&&+\rho_{12}^{\mathrm{QCD-A}}\left(s,u,Q^2\right)+\left[m_{\mathcal{B}_{2}^{*-}}+m_{\mathcal{B}_{1}^{-}}-m_{\mathcal{B}_{2}^{*+}}\right]\rho_{11}^{\mathrm{QCD-A}}\left(s,u,Q^2\right)
+\left[m_{\mathcal{B}_{2}^{*-}}+m_{\mathcal{B}_{1}^{-}}-m_{\mathcal{B}_{1}^{+}}\right]\rho_{10}^{\mathrm{QCD-A}}\left(s,u,Q^2\right)\Big\} \\ \notag
G_{3}^{++}(Q^2) &&=
-\frac{m_{\mathcal{B}_{1}^{+}}m_{\mathcal{B}_{2}^{*+}}e^{\frac{m_{\mathcal{B}_{1}^{+}}^{2}}{M_{1}^{2}}+\frac{m_{\mathcal{B}_{2}^{*+}}^{2}}{M_{2}^{2}}}}
{\left(m_{\mathcal{B}_{2}^{*+}}+m_{\mathcal{B}_{2}^{*-}}\right)\left(m_{\mathcal{B}_{1}^{+}}+m_{\mathcal{B}_{1}^{-}}\right)\lambda_{\mathcal{B}_{2}^{*+}}\lambda_{\mathcal{B}_{1}^{+}}}
\int^{s_{0}}_{s_{min}}ds\int^{u_{0}}_{u_{min}}du\Big\{\left[Q^{2}-m_{\mathcal{B}_{2}^{*-}}m_{\mathcal{B}_{1}^{-}}+m_{\mathcal{B}_{1}^{+}}m_{\mathcal{B}_{1}^{-}}
+m_{\mathcal{B}_{2}^{*+}}m_{\mathcal{B}_{2}^{*-}}\right]\rho_{13}^{\mathrm{QCD-A}}\left(s,u,Q^2\right)
\\ \notag
&&+\rho_{16}^{\mathrm{QCD-A}}\left(s,u,Q^2\right)+\left[m_{\mathcal{B}_{1}^{-}}-m_{\mathcal{B}_{1}^{+}}+m_{\mathcal{B}_{2}^{*-}}\right]\rho_{14}^{\mathrm{QCD-A}}\left(s,u,Q^2\right)
+\left[m_{\mathcal{B}_{2}^{*-}}-m_{\mathcal{B}_{2}^{*+}}+m_{\mathcal{B}_{1}^{-}}\right]\rho_{15}^{\mathrm{QCD-A}}\left(s,u,Q^2\right) \\ \notag
&&+2\rho_{3}^{\mathrm{QCD-A}}\left(s,u,Q^2\right)+2\left(m_{\mathcal{B}_{2}^{*+}}-m_{\mathcal{B}_{1}^{-}}-m_{\mathcal{B}_{2}^{*-}}\right)\rho_{4}^{\mathrm{QCD-A}}\Big\} \\ \notag
G_{4}^{++}(Q^2) &&=
-\frac{e^{\frac{m_{\mathcal{B}_{1}^{+}}^{2}}{M_{1}^{2}}+\frac{m_{\mathcal{B}_{2}^{*+}}^{2}}{M_{2}^{2}}}}
{\left(m_{\mathcal{B}_{2}^{*+}}+m_{\mathcal{B}_{2}^{*-}}\right)\left(m_{\mathcal{B}_{1}^{+}}+m_{\mathcal{B}_{1}^{-}}\right)\lambda_{\mathcal{B}_{2}^{*+}}\lambda_{\mathcal{B}_{1}^{+}}}
\int^{s_{0}}_{s_{min}}ds\int^{u_{0}}_{u_{min}}du\Big\{\left[Q^{2}+m_{\mathcal{B}_{2}^{*+}}m_{\mathcal{B}_{2}^{*-}}-m_{\mathcal{B}_{2}^{*-}}m_{\mathcal{B}_{1}^{-}}+m_{\mathcal{B}_{1}^{+}}m_{\mathcal{B}_{1}^{-}}
\right]\rho_{5}^{\mathrm{QCD-A}}\left(s,u,Q^2\right) \\
&&+\rho_{8}^{\mathrm{QCD-A}}\left(s,u,Q^2\right)+\left[m_{\mathcal{B}_{1}^{-}}-m_{\mathcal{B}_{1}^{+}}+m_{\mathcal{B}_{2}^{*-}}\right]\rho_{6}^{\mathrm{QCD-A}}\left(s,u,Q^2\right)
+\left[m_{\mathcal{B}_{2}^{*-}}+m_{\mathcal{B}_{1}^{-}}-m_{\mathcal{B}_{2}^{*+}}\right]\rho_{7}^{\mathrm{QCD-A}}\left(s,u,Q^2\right) \Big\}
\end{eqnarray}

\section{The integral formulas for two and three Dirac delta functions}\label{Sec:AppB}
The integral formulas of two Dirac delta functions can be expressed as,
\begin{eqnarray}\label{eq:B1}
\notag
\int {{d^4}k\delta ({k^2} - m_1^2)\delta [{{(q' - k)}^2} - m_4^2]} &&= \frac{{\pi \sqrt {\lambda (r,m_1^2,m_4^2)} }}{{2r}}\\ \notag
\int {{d^4}k\delta ({k^2} - m_1^2)\delta [{{(q' - k)}^2} - m_4^2]} {k_\mu } &&= \frac{{\pi \sqrt {\lambda (r,m_1^2,m_4^2)} }}{{2r}}\frac{{r + m_1^2 - m_4^2}}{{2r}}q{'_\mu }\\
\int {{d^4}k\delta ({k^2} - m_1^2)\delta [{{(q' - k)}^2} - m_4^2]} {k_\mu }{k_\nu } &&= \frac{{\pi \sqrt {\lambda (r,m_1^2,m_4^2)} }}{{2r}}\left[ { - \frac{{\lambda (r,m_1^2,m_4^2)}}{{12r}}{g_{\mu \nu }} + \left( {m_1^2 + \frac{{\lambda (r,m_1^2,m_4^2)}}{{3r}}} \right)\frac{{q{'_\mu }q{'_\nu }}}{{q{'^2}}}} \right]
\end{eqnarray}
where, $r=q'^2$ and $\lambda(a,b,c)=a^2+b^2+c^2-2(ab+ac+bc)$ is the triangle function.

The integral formulas of three Dirac delta functions can be given as follows,
\begin{eqnarray}\label{eq:B2}
\notag
&&\int {{d^4}k} \delta ({k^2} - m_3^2)\delta [{(k + p-p')^2} - m_2^2]\delta [{(p' - k)^2} - m^2] = \frac{\pi }{{2\sqrt {\lambda (s,u,{q^2})} }}\\ \notag
&&\int {{d^4}k} \delta ({k^2} - m_3^2)\delta [{(k + p-p')^2} - m_2^2]\delta [{(p' - k)^2} - m^2]{k_\mu } = \frac{\pi }{{2\sqrt {\lambda (s,u,{q^2})} }}({\alpha _1}{p_\mu } + {\beta _1}p{'_\mu })\\
\notag
&&\int {{d^4}k} \delta ({k^2} - m_3^2)\delta [{(k + p-p')^2} - m_2^2]\delta [{(p' - k)^2} - m^2]{k_\mu }{k_\nu } = \\ \notag
&&\frac{\pi }{{2\sqrt {\lambda (s,u,{q^2})} }}[{\alpha _2}{g_{\mu \nu }} + {\beta _2}{p_\mu }{p_\nu } + {\gamma _2}(p{'_\mu }{p_\nu } + {p_\mu }p{'_\nu }) + {\delta _2}p{'_\mu }p{'_\nu }] \\ \notag
&&\int {{d^4}k} \delta ({k^2} - m_3^2)\delta [{(k + p-p')^2} - m_2^2]\delta [{(p' - k)^2} - m^2]{k_\mu }{k_\nu }{k_\lambda } = \\ \notag
&&\frac{\pi }{{2\sqrt {\lambda (s,u,{q^2})} }}[{\alpha _3}({p_{\mu}g_{ \nu\lambda }}+{p_{\nu}g_{ \mu\lambda }}+{p_{\lambda}g_{ \mu\nu }}) + {\lambda _3}({p{'_\mu}g_{ \nu\lambda }}+{p{'_\nu}g_{ \mu\lambda }}+{p{'_\lambda}g_{ \mu\nu }}) + {\gamma _3}p_{\mu }{p_\nu }{p_\lambda } \\
&&+ {\rho _3}(p{'_\mu}p_{\nu }p_{\lambda }+p{'_\nu}p_{\mu }p_{\lambda }+p{'_\lambda}p_{\mu }p_{\nu })+{\delta _3}(p_{\mu}p{'_\nu }p{'_\lambda }+p_{\nu}p{'_\mu }p{'_\lambda }+p_{\lambda}p{'_\nu }p{'_\mu })+\kappa_{3}p{'_\mu}p{'_\nu}p{'_\beta\lambda}]
\end{eqnarray}
where $s=p^2$, $u=p'^2$, $q=p-p'$,
\begin{eqnarray}\label{eq:B3}
\notag
{\alpha _1}&&= \frac{m^2 \left(q^2-s+u\right)+u \left(-2 \text{m}_{2}^2+q^2+s-u\right)+\text{m}_{3}^2 \left(-q^2+s+u\right)}{\lambda (s,u,{q^2})}\\
{\beta _1} &&=\frac{m^2 \left(q^2+s-u\right)+\text{m}_{2}^2 \left(-q^2+s+u\right)-2 \text{m}_{3}^2 s+q^4-q^2 s-2 q^2 u-s u+u^2}{\lambda (s,u,{q^2})}
\end{eqnarray}
\begin{eqnarray}\label{eq:B4}
\notag
{\alpha _2} &&=\frac{1}{2 \lambda (s,u,{q^2})}\Bigg\{m^4 q^2-m^2 \left[\text{m}_{2}^2 \left(q^2-s+u\right)+\text{m}_{3}^2 \left(q^2+s-u\right)+q^2 \left(-q^2+s+u\right)\right]+\text{m}_{2}^4 u\\ \notag
&&+\text{m}_{2}^2 \left[\text{m}_{3}^2 \left(q^2-s-u\right)+u \left(-q^2-s+u\right)\right]+s \left[\text{m}_{3}^4-\text{m}_{3}^2 \left(q^2-s+u\right)+q^2 u\right]\Bigg\}\\
\notag
{\beta _2} &&=\frac{1}{\lambda (s,u,{q^2})^2}
\Bigg\{m^4 \left[q^4-2 q^2 (s-2 u)+(s-u)^2\right]-2 m^2 \Big[u \left(3 \text{m}_{2}^2 (q^2-s+u)-2 q^4+q^2 (s+u)+(s-u)^2\right)\\ \notag
&&+\text{m}_{3}^2 \left(q^4+q^2 (u-2 s)+s^2+s u-2 u^2\right)\Big]-2 \text{m}_{3}^2 u \left[3 \text{m}_{2}^2 (-q^2+s+u)+q^4+q^2 (s-2 u)-2 s^2+s u+u^2\right]\\ \notag
&&+u^2 \left[6 \text{m}_{2}^4-6 \text{m}_{2}^2 (q^2+s-u)+q^4+q^2 (4 s-2 u)+(s-u)^2\right]+\text{m}_{3}^4 \left[q^4-2 q^2 (s+u)+s^2+4 s u+u^2\right]\Bigg\}\\
\notag
{\gamma _2} &&=-\frac{1}{\lambda (s,u,{q^2})} 2 \Bigg\{(-m^2+\text{m}_{3}^2+u) (-m^2+\text{m}_{2}^2-q^2+u)-\frac{1}{\lambda (s,u,{q^2})}3 (q^2-s-u) \Big[m^4 q^2-m^2 \Big(\text{m}_{2}^2 (q^2-s+u)\\ \notag
&&+\text{m}_{3}^2 (q^2+s-u)+q^2 (-q^2+s+u)\Big)+\text{m}_{2}^4 u+\text{m}_{2}^2 \Big(\text{m}_{3}^2 (q^2-s-u)+u (-q^2-s+u)\Big)+s \Big(\text{m}_{3}^4-\text{m}_{3}^2 (q^2-s+u)+q^2 u\Big)\Big]\\ \notag
&&-2 \text{m}_{3}^2 (-q^2+s+u)\Bigg\}\\
\notag
{\delta _2} &&=\frac{1}{\lambda (s,u,{q^2})^2} \Bigg\{m^4 \Big[q^4+q^2 (4 s-2 u)+(s-u)^2\Big]-2 m^2 \Big[\text{m}_{2}^2 \Big(q^4+q^2 (s-2 u)-2 s^2+s u+u^2\Big)+q^2 (3 \text{m}_{3}^2 s+2 s^2+3 s u-3 u^2)\\ \notag
&&+(s-u) \Big(3 \text{m}_{3}^2 s+u (s-u)\Big)-q^6-q^4 (s-3 u)\Big]+\text{m}_{2}^4 \Big(q^4-2 q^2 (s+u)+s^2+4 s u+u^2\Big)-2 \text{m}_{2}^2 \Big(q^2 (-3 \text{m}_{3}^2 s+s^2+3 s u+3 u^2)\\ \notag
&&+3 \text{m}_{3}^2 s (s+u)+q^6-q^4 (2 s+3 u)-u (-2 s^2+s u+u^2)\Big)+6 \text{m}_{3}^4 s^2-4 \text{m}_{3}^2 q^4 s+2 \text{m}_{3}^2 q^2 s^2+8 \text{m}_{3}^2 q^2 s u+2 \text{m}_{3}^2 s^3+2 \text{m}_{3}^2 s^2 u\\
&&-4 \text{m}_{3}^2 s u^2+q^8-2 q^6 s-4 q^6 u+q^4 s^2+2 q^4 s u+6 q^4 u^2+4 q^2 s^2 u+2 q^2 s u^2-4 q^2 u^3+s^2 u^2-2 s u^3+u^4\Bigg\}
\end{eqnarray}
The expressions of coefficients $\alpha_{3}\sim\kappa_{3}$ are too complicated to give out explicitly. In addition, a geometric constraint will be introduced in the integration of three Dirac delta functions, and it has the following form,
\begin{eqnarray}\label{eq:B5}
- 1 \le \cos \theta  = \frac{{(u - {q^2} + m_2^2 - m^2)(s + u - {q^2}) + 2s(m^2-u - m_3^2)}}{{\sqrt {{{(u - {q^2} + m_2^2 - m^2)}^2} - 4sm_3^2} \sqrt {\lambda (s,u,q^2)} }} \le 1
\end{eqnarray}
\end{widetext}
\end{document}